\PassOptionsToPackage{table,xcdraw}{xcolor}
\documentclass[format=acmsmall, review=false, screen=true, authorversion]{acmart}
\AtBeginDocument{%
  \providecommand\BibTeX{{%
    \normalfont B\kern-0.5em{\scshape i\kern-0.25em b}\kern-0.8em\TeX}}}

\newcommand{\ie}{i.e.}
\newcommand{\eg}{e.g.}
\newcommand{\etal}{\textit{et al}.}

\usepackage{tablefootnote}

\usepackage{moreverb}
\usepackage{xspace}
\usepackage{multirow}
\usepackage{xcolor}
\usepackage[normalem]{ulem}
\usepackage{multirow}
\usepackage{amsmath}
\usepackage{bm}

\newcommand{\bert}{\textsc{\small{BERT}}\xspace}
\newcommand{\bertrec}{\textsc{\small{BERT4Rec}}\xspace}
\newcommand{\sasrec}{\textsc{\small{SASRec}}\xspace}
\newcommand{\grurec}{\textsc{\small{GRU4Rec}}\xspace}
\newcommand{\sknn}{\textsc{\small{SKNN}}\xspace}
\newcommand{\ssknn}{\textsc{\small{S\_SKNN}}\xspace}
\newcommand{\vsknn}{\textsc{\small{V\_SKNN}}\xspace}
\newcommand{\zesrec}{\textsc{\small{ZESRec}}\xspace}
\newcommand{\vqrec}{\textsc{\small{VQ-Rec}}\xspace}
\newcommand{\gptrec}{\textsc{\small{GPTRec}}\xspace}
\newcommand{\pfive}{\textsc{\small{P5}}\xspace}
\newcommand{\once}{\textsc{\small{ONCE}}\xspace}
\newcommand{\naml}{\textsc{\small{NAML}}\xspace}
\newcommand{\nrms}{\textsc{\small{NRMS}}\xspace}
\newcommand{\fastformer}{\textsc{\small{Fastformer}}\xspace}
\newcommand{\mmrec}{\textsc{\small{MM-Rec}}\xspace}
\newcommand{\transrec}{\textsc{\small{TransRec}}\xspace}
\newcommand{\unitrec}{\textsc{\small{UniTRec}}\xspace}
\newcommand{\promptfornr}{\textsc{\small{Prompt4NR}}\xspace}
\newcommand{\bookgpt}{\textsc{\small{BookGPT}}\xspace}
\newcommand{\tallrec}{\textsc{\small{TALLRec}}\xspace}
\newcommand{\pbnr}{\textsc{\small{PBNR}}\xspace}
\newcommand{\gptforrec}{\textsc{\small{GPT4Rec}}\xspace}
\newcommand{\palr}{\textsc{\small{PALR}}\xspace}
\newcommand{\youtubednn}{\textsc{\small{YouTubeDNN}}\xspace}
\newcommand{\chatrec}{\textsc{\small{Chat-Rec}}\xspace}
\newcommand{\srec}{\textsc{\small{S3-Rec}}\xspace}
\newcommand{\llamaGenItemModel}{\textsc{\small{LlamaSeqPromptGenItem}}\xspace}

\newcommand{\dhr}{Delivery Hero\xspace}

\newcommand{\embeddingModel}{\textsc{\small{LLMSeqSim}}\xspace}
\newcommand{\promptModel}{\textsc{\small{LLMSeqPrompt}}\xspace}
\newcommand{\bertmodel}{\textsc{\small{LLM2BERT4Rec}}\xspace}
\newcommand{\sasrecmodel}{\textsc{\small{LLM2SASRec}}\xspace}
\newcommand{\grurecmodel}{\textsc{\small{LLM2GRU4Rec}}\xspace}
\newcommand{\sknnmodel}{\textsc{\small{SKNNEmb}}\xspace}
\newcommand{\seqmodel}{\textsc{\small{LLM2Sequential}}\xspace}

\newcommand{\popHybrid}{\textsc{\small{\embeddingModel\&Sequential}}\xspace}
\newcommand{\embeddingPop}{\textsc{\small{\embeddingModel\&Popularity}}\xspace}
\newcommand{\promptGenItemModel}{\textsc{\small{LLMSeqPromptGenItem}}\xspace}
\newcommand{\promptGenListModel}{\textsc{\small{LLMSeqPromptGenList}}\xspace}
\newcommand{\promptClassModel}{\textsc{\small{LLMSeqPromptClass}}\xspace}
\newcommand{\promptRankModel}{\textsc{\small{LLMSeqPromptRank}}\xspace}
\newcommand{\bertbertrec}{\textsc{\small{BERT2BERT4Rec}}\xspace}
\newcommand{\bertsasrec}{\textsc{\small{BERT2SASRec}}\xspace}
\newcommand{\bertgrurec}{\textsc{\small{BERT2GRU4Rec}}\xspace}

\mathchardef\mhyphen="2D % Define a "math hyphen"
\newcommand\pcore{\mathop{p\mhyphen \mathrm{core}}}

\rowcolors{2}{gray!25}{white}

\begin{document}

\title{Improving Sequential Recommendations with LLMs}
%% Of note is the shared affiliation of the first two authors, and the
%% "authornote" and "authornotemark" commands
%% used to denote shared contribution to the research.
\author{Artun Boz}
\affiliation{%
  \institution{Individual contributor}
  \city{Delft}
  \country{The Netherlands}
}
\authornote{Part of the author's work was carried out while at Delivery Hero Research.}

\author{Wouter Zorgdrager\small{*}}
\affiliation{%
  \institution{Individual contributor}
  \city{Delft}
  \country{The Netherlands}
}

\author{Zoe Kotti\small{*}}
\affiliation{%
  \institution{Athens University of Economics \& Business}
  \city{Athens}
  \country{Greece}
}

\author{Jesse Harte\small{*}}
\affiliation{%
  \institution{Delft University of Technology}
  \city{Delft}
  \country{The Netherlands}
}

\author{Panos Louridas}
\affiliation{%
  \institution{Athens University of Economics \& Business}
  \city{Athens}
  \country{Greece}
}

\author{Vassilios Karakoidas}
\affiliation{%
  \institution{Delivery Hero}
  \city{Berlin}
  \country{Germany}
}

\author{Dietmar Jannach}
\affiliation{%
  \institution{University of Klagenfurt}
  \city{Klagenfurt}
  \country{Austria}
}

\author{Marios Fragkoulis\small{*}}
\affiliation{%
  \institution{Delft University of Technology}
  \city{Delft}
  \country{The Netherlands}
}

\renewcommand{\shortauthors}{Boz et al.}

\begin{abstract}
  The sequential recommendation problem has % %\zoe{recently}
  attracted considerable
  research attention in the past few years, leading to the rise of
  numerous recommendation models. In this work, we explore how Large
  Language Models (LLMs), which are nowadays introducing disruptive
  effects in many AI-based applications, can be used to build or
  improve sequential recommendation approaches. Specifically, we
  design three orthogonal approaches and hybrids of those to leverage
  the power of LLMs in different ways. In addition, we investigate the potential of each approach
  by focusing on its comprising technical aspects and
  determining an array of alternative choices for each one. We
  conduct extensive experiments on three datasets and explore a
  large variety of configurations, including different language models and baseline
  recommendation models,
  to obtain a comprehensive picture of the performance of each approach.

  Among other observations, we
  highlight that initializing state-of-the-art sequential
  recommendation models such as BERT4Rec or SASRec with embeddings
  obtained from an LLM can lead to substantial performance gains in
  terms of accuracy. Furthermore, we find that fine-tuning an LLM for recommendation tasks
  enables it to learn not only the tasks, but also concepts of a domain to some extent.
  We also show that fine-tuning OpenAI GPT leads to considerably better performance than fine-tuning Google PaLM 2.
  Overall, our extensive experiments indicate a huge potential value
  of leveraging LLMs in future recommendation approaches.
We publicly share the code and data of our experiments to ensure reproducibility.%\dietmar{ fix URL}
\footnote{\url{https://github.com/dh-r/LLM-Sequential-Recommendation}}
\end{abstract}

%%
%% The code below is generated by the tool at http://dl.acm.org/ccs.cfm.
%% Please copy and paste the code instead of the example below.
%%

%\maketitle

%%
%% Keywords. The author(s) should pick words that accurately describe
%% the work being presented. Separate the keywords with commas.

%%
%% This command processes the author and affiliation and title
%% information and builds the first part of the formatted document.
\maketitle

%%% ==================================================
\section{Introduction}
\label{sec:introduction}
%%% ==================================================
The sequential recommendation problem considers a sequence of past user interactions in order to predict the next user interest or action. 
The problem's wide application in many popular domains, such as
next-purchase prediction~\cite{Jannach2015adaptation}, next-track
music recommendation~\cite{Chen2012playlist}, and next
Point-of-Interest suggestions for tourism~\cite{Lim2015personalized},
has sparked substantial research interest in the recent years. This
had led to the inception of novel algorithmic
approaches~\cite{Hidas2016session,Ludewig2018,Kang2018selfattentive,Sun2019BERT4Rec},
including approaches that utilize side information about the items,
such as an item's category~\cite{xu2022category,lai2022attribute}.

Lately, the advent of Large Language Models (LLM), which in an
\emph{auto-regressive mode} repeatedly complete a sequence until they
fulfil a prompt task~\cite{radford2018improving}, 
has opened a fruitful research direction for
sequential recommendation. In fact, LLMs based on Generative
Pretrained Transformers~\cite{radford2019language} are only the
latest of a stream of innovations in Natural Language Processing (NLP) that
inspired the birth of new sequential recommendation models. Notable
examples are~\grurec~\cite{Hidas2016session},
\sasrec~\cite{Kang2018selfattentive}, and
\bertrec~\cite{Sun2019BERT4Rec}, which were influenced by
the Gated Recurrent Unit (GRU)~\cite{cho2014learning}, the transformer
architecture~\cite{Vaswani2017attention}, and
\bert~\cite{Devlin2019BERT}, respectively. The key contribution of
LLMs to the sequential recommendation problem is a semantic model of
real world concepts and their representation in natural language
sequences. 
This consideration is neglected by typical ID-based sequential recommendation models~\cite{Kang2018selfattentive, Sun2019BERT4Rec}.

The rapidly growing interest for involving LLMs in recommender systems
clusters around cross-domain~\cite{Zhang2021language, hou2023large},
domain-specific~\cite{bao2023tallrec, zhang2023prompt}, and
multi-modal~\cite{wu2022mmrec, wang2022transrec} recommendations.
Within the scope of recommendation approaches, LLMs are involved
either by providing item embeddings~\cite{Hou2023learning} or by
responding to a recommendation task conveyed in a prompt, which may
contain zero~\cite{Ding2022zero} or a few
examples~\cite{Geng2022Recommendation} to enrich the context of the
task available to an LLM. Alternatively, an LLM can be fine-tuned to
learn how to perform better for a specific recommendation
task~\cite{hou2022towards}. Regarding the nature of recommendation
tasks posed to LLMs, they are typically used to
generate~\cite{li2023gpt4rec} or rerank~\cite{hou2023large}
recommendations. Altogether, the current state of research is
summarized in the following surveys~\cite{lin2023recommender,
  Wu2023survey}.

In this paper, we build on the ideas developed in these early works on involving LLMs in recommender systems, and
we design and research the potential of three
orthogonal approaches and hybrids of those that leverage LLMs for the particular problem setting of sequential recommendations. We decompose each approach into its constituent technical aspects, identify or devise alternatives for
each aspect, and investigate the impact of choices to the approach's
performance. Finally, we thoroughly tune and evaluate all approaches
under the same testbed across three datasets of different domains and
characteristics.
These datasets are Amazon Beauty from the e-commerce beauty domain, a proprietary \dhr dataset from the e-commerce darkstore domain, and Steam from the gaming domain.

In the first approach (\embeddingModel), 
we retrieve a
semantically-rich embedding from an existing LLM for each item in a
session, optionally reduce the embedding
to a target number of dimensions, and
then compute an aggregate session embedding. We use the latter to recommend catalog
products with a similar embedding. In this first approach we investigate the following four technical aspects: the
choice of LLM to retrieve embeddings from, dimensionality reduction
methods, target embedding dimensions, and session embedding
computation strategies. In the second approach (\promptModel), we
fine-tune an LLM with dataset-specific information in the form of
prompt-completion pairs and ask the model to produce next item
recommendations for a set of test prompts.
We explore four different aspects of LLMs, namely models, versions, parameters, and task specifications. Our third approach
(\seqmodel) consists of enhancing existing sequential models with item
embeddings obtained from an LLM. Here, we examine LLM embedding
models, dimensionality reduction methods, and sequential models.
Finally, we create hybrid recommendation approaches based on the aforementioned ones.

Our work results in the following contributions and insights.\footnote{The present paper substantially extends our previous work on leveraging LLMs for the sequential recommendation problem~\cite{harte2023leveraging}. We detail our introduced extensions in Section~\ref{sec:background}.}

\begin{itemize}
\item Building on the growing literature regarding the use of LLMs in recommender systems, we design and research three orthogonal methods and two hybrid   approaches of leveraging LLMs for sequential recommendation. In
  particular, we investigate the technical aspects of each method,
  determine alternatives for each, thoroughly tune them, and
  evaluate their impact.

\item Experiments on three datasets, including a proprietary
  real-world dataset from \dhr, reveal that LLM embeddings boost the
  performance of different classes of sequential models across all
  datasets. Specifically, \sasrecmodel and \bertmodel, which are implementations of the \seqmodel approach, top the
  leaderboard increasing NDCG@20 on average by 45\% on the Amazon Beauty dataset and 9\%
  on the \dhr dataset. Beyond accuracy metrics, \sasrecmodel
  almost doubles catalog coverage and substantially increases serendipity by
  21\% over \sasrec across all datasets.

\item Both OpenAI GPT and Google PaLM fine-tuned models (\promptModel) perform significantly better than their counterpart base models across tasks marking 16\% average improvement in terms of NDCG@20. Most importantly, the fine-tuned models exhibit as low as half the number of hallucinations and noticeably higher semantic similarity to actual recommendation options compared to the base models indicating that fine-tuned models can learn not only tasks, but also concepts of a domain to some extent. Finally, \texttt{\small{GPT 3.5 turbo}} considerably outperforms \texttt{\small{PaLM 2 bison}} by more than 78\% on average in terms of NDCG@20 across fine-tuning tasks and datasets.

\item In one of the datasets, Amazon Beauty, the semantic item recommendation model via LLM embeddings (\embeddingModel) takes the third place and achieves the best MRR score, while a fine-tuned model (\promptModel) outperforms \grurec and \sasrec. Both results support the potential of LLM-based models for datasets with specific characteristics.

\end{itemize}

Overall, the main contribution of our work is an in-depth analysis of the potential of various previously-explored and novel ways of leveraging LLMs for sequential recommendation problems. Our comprehensive analyses, which involve different datasets, three LLMs, and a set of orthogonal technical approaches reveal that substantial performance improvements can indeed be obtained by involving LLMs for sequential recommendations.

The rest of the paper is organized as follows.
Section~\ref{sec:background} presents the background and related work.
Next,
Sections~\ref{sec:embeddings},~\ref{sub:fine-tuned-LLM}, ~\ref{sec:sequential},
and~\ref{sec:hybrids} present the researched recommendation
approaches.
Then, Section~\ref{sec:experiments} details the
experiments and, finally, Section~\ref{sec:conclusions} conveys the
conclusions of our work.

%%%%%
%%% ==================================================
\section{Background \& Related Work}
\label{sec:background}

The recent developments in LLMs have taken the world by surprise.
Models like OpenAI GPT~\cite{Brown2020Language}, Google
BERT~\cite{Devlin2019BERT}, and Meta LLaMA~\cite{touvron2023llama},
which employ deep transformer architectures, demonstrate how
innovations in NLP 
can reshape and advance mainstream online activities, such as search,
shopping, and customer care. Inevitably, research in recommender
systems is significantly impacted by the developments in the area of
LLMs as well.

According to recent surveys~\cite{lin2023recommender, Wu2023survey},
LLMs are mainly utilized for recommendation problems in two ways: by
providing embeddings that can be used to initialize existing
recommendation models~\cite{Wu2021Empowering, Zhang2021unbert,
  Liu21pretrained}, and by producing recommendations leveraging their
inherent knowledge encoding~\cite{kang2023do, Hegselmann23tabllm,
  bao2023tallrec}. LLMs as recommendation models can provide
recommendations given (i) only a task specification (zero-shot), (ii)
one or few examples given inline to the prompt of the task (few-shot),
or (iii) after fine-tuning the model's weights to the task at hand
given a set of training examples~\cite{Brown2020Language}. This
incremental training process deviates from typical recommendation
models, which have to be trained from zero on domain data.
% \zoe{from scratch}
% \marios{equivalent (matter of taste ;-)}
In fact, LLMs show early indications of % easy
adaptability to different recommendation domains with modest
fine-tuning~\cite{hou2022towards, Hou2023learning}. Finally, LLMs have
so far been used for a number of recommendation tasks, such as rating
prediction~\cite{li2023text}, item generation~\cite{li2023gpt4rec},
and reranking~\cite{hou2023large} across domains (\eg,
news~\cite{Wu2021Empowering}, information
retrieval~\cite{Liu21pretrained}).

In this work we explore the potential of using LLMs for sequential
recommendation problems~\cite{jannach2020research}. In these problems,
we consider as input a sequence of user interactions (\ie a session)
$S^u = (S^u_1, S^u_2, \ldots, S^u_n)$, where $u$ is a user, $n$ is the
length of the sequence and $S^u_i$ are individual items. The aim is to
predict the next interaction of the given sequence. Besides the recent sequential recommendation models mentioned in the
introduction~\cite{Hidas2016session, Kang2018selfattentive,
  Sun2019BERT4Rec}, \srec~\cite{CIKM2020-S3Rec} pre-trains a transformer architecture enhanced with self-supervised signals aiming to learn better data representations using the mutual information maximization principle. 
In earlier works the sequential recommendation
problem has been modelled as a Markov
Chain~\cite{Garcin2013personalized} or a Markov Decision
Process~\cite{Shani2005MDP}. Neighborhood-based approaches, such as
SKNN~\cite{Jannach2017when}, have also been proposed.

Related pieces of work in the sequential recommendation problem focus
on cross-domain (Section~\ref{sec:cross-domain-llms}), domain-specific
(Section~\ref{sec:domain-specific-llms}), and multi-modal
(Section~\ref{sec:multimodal}) recommendations.

\subsection{Cross-domain LLM-based Sequential Recommendation Models}
\label{sec:cross-domain-llms}

Early research work regarding LLMs for sequential recommendation
problems showed mixed results for zero-shot and fine-tuned
recommendations~\cite{Zhang2021language, Geng2022Recommendation,
  Ding2022zero, liu2023chatgpt, hou2023large, Hou2023learning,
  wang2023zeroshot, sileo2022zeroshot}. The first deep zero-shot
generative recommender model named \zesrec~\cite{Ding2022zero}
comprises four unique properties: \emph{cold users} (\ie, no
overlapping users between training and test data); \emph{cold items}
(\ie, no overlapping items between training and test data);
\emph{domain gap} (\ie, training and test data originate from different
domains); and \emph{no access to target data} (\ie, target data are
only available at inference time). \zesrec outperforms the zero-shot
embedding-KNN and random baselines substantially, and proves
beneficial for data-scarce startups and early-stage products.
In~\cite{wang2023zeroshot} and~\cite{Zhang2021language} both
GPT~\cite{radford2018improving} and \bert~\cite{Devlin2019BERT}
perform better than a random baseline in zero-shot sequential
recommendations, while \grurec~\cite{Hidas2016session} outperforms the
fine-tuned versions of the LLMs. \grurec and other baselines also
outperform \zesrec~\cite{Ding2022zero} for zero-shot recommendations
across domains. Conversely, \pfive~\cite{Geng2022Recommendation},
which can produce zero-shot or few-shot recommendations without
fine-tuning,  
has been shown to outperform a number of sequential recommendation
models and ChatGPT---the least performant of all
models~\cite{liu2023chatgpt}.

The very recent \vqrec model~\cite{Hou2023learning} employs a
transformer architecture and applies a novel representation scheme to
embeddings retrieved from \bert to adapt to new domains. \vqrec
outperforms a number of sequential recommendation models across
datasets of different
domains. 
\chatrec~\cite{gao2023chatrec} converts user profiles and past
interactions into prompts to build a conversational recommender system
based on ChatGPT. The system is effective in learning user preferences
and transferring them to different products, potentially allowing
cross-domain recommendations.
Finally, in both~\cite{Hou2023learning} and~\cite{yuan2023where} \sasrec with LLM embeddings is shown to improve over \sasrec.

To explore the performance limits of text-based collaborative
filtering with LLMs, Li \etal~\cite{li2023exploring} conduct a series
of experiments contrasting the former with the dominant
identifier-based approach and the trending prompt-based recommendation
using ChatGPT. Interestingly, ChatGPT seems inferior than text-based
collaborative filtering, while the simple identifier-based approach
remains highly competitive in the warm item recommendation setting.
However, in the book recommendation tasks (\ie, book rating, user
rating, and book summary recommendation), a ChatGPT-like system
(\bookgpt) developed by Li \etal~\cite{li2023bookgpt} achieves
promising results, especially in zero-shot or one-shot learning,
compared to the classic book recommendation algorithms. Another
investigation on the capabilities of ChatGPT in recommender
systems~\cite{dai2023uncovering} highlights that the model performs
better with list-wise ranking, compared to point-wise and pair-wise.
Overall, LLM-based recommenders surpass popularity-based recommenders
in the movie, book, and music domains, but not so in the news domain.

\subsection{Domain-specific LLM-based Sequential Recommendation Models}
\label{sec:domain-specific-llms}

In the \emph{movie, book, and e-commerce} domains,
the lightweight \tallrec framework proposed by Bao \etal~\cite{bao2023tallrec}
aims to adapt LLMs for recommendations
by structuring the recommendations as instructions, and
tuning the LLMs through an instruction-tuning process.
Experiments using the LLaMA model~\cite{touvron2023llama}
demonstrate that \tallrec significantly advances the LLM capabilities, and
is highly efficient even on a single conventional GPU.
% Movie, E-commerce
Similarly, \palr~\cite{yang2023palr} employs user/item interactions
for candidate retrieval, and a fine-tuned LLM-based ranking model to
produce recommendations. Using LLaMA for the LLM, and compared to
representative baselines including \grurec~\cite{Hidas2016session}
and \sasrec~\cite{Kang2018selfattentive}, \palr appears more effective
in various sequential recommendation tasks.

Furthermore, \gptrec~\cite{petrov2023generative},
which is based on the GPT-2 architecture~\cite{radford2019language},
employs a novel SVD tokenization algorithm and a Next-K recommendation strategy
to generate sequential recommendations,
accounting for already recommended items.
\gptrec matches the quality of \sasrec~\cite{Kang2018selfattentive} but is more GPU-memory efficient
due to its sub-item tokenization. Inspired by search engines, \gptforrec~\cite{li2023gpt4rec} generates
a set of search queries based on the item titles from a user's
history, and then recommends items by searching these queries using
beam search. Using GPT-2~\cite{radford2019language} as a language
model to generate queries that are used in a search engine using
BM25~\cite{robertson2009the} as its score function, \gptforrec
outperforms state-of-the-art models including
\youtubednn~\cite{covington2016deep} and
\bertrec~\cite{Sun2019BERT4Rec}. In addition, multi-query generation
with beam search increases the diversity of recommended items and the
coverage of a user's interests.

In the \emph{news} domain, Zhang and Wang~\cite{zhang2023prompt} have
recently released a framework for prompt learning (\promptfornr) to
predict whether a user would click a candidate news item. \promptfornr
slightly outperforms BERT-based models, reflecting the usefulness of
the prompt learning approach for embedding knowledge in the
pre-trained BERT models employed for news recommendation. Similarly,
the prompt-based news recommender system (\pbnr) by Li
\etal~\cite{li2023pbnr} leverages the text-to-text T5
model~\cite{Raffel2020exploring} for personalized recommendations.
\pbnr is very accurate even with varying lengths of past user
interactions, adaptable to new data, and can also satisfy user
requirements through human-computer interaction.
Finally, Li \etal~investigate ChatGPT's performance in news
recommendation~\cite{li2023preliminary} with respect to personalized recommendations, fairness, and fake news detection.

To enhance content-based recommendation, Liu \etal~\cite{liu2023once}
propose a framework named \once which combines the deep layers of
open-source LLMs (\ie, LLaMA~\cite{touvron2023llama}) as content
encoders to improve the representation embeddings, and the prompting
techniques of closed-source LLMs (\ie, ChatGPT~\cite{liu2023chatgpt})
to enrich the training token data. \once shows a substantial
improvement of 19.32\% compared to the state-of-the-art
(\naml~\cite{wu2019neural}, \nrms~\cite{an2019neural},
\fastformer~\cite{wu2019bneural}), showcasing the synergistic
relationship between fine-tuning on the open-source LLMs and prompting
on the closed-source LLMs.
\unitrec, a unified text-to-text transformer and joint contrastive
learning framework for text-based recommendation is proposed by Mao
\etal~\cite{mao2023unitrec} to better model two-level contexts of user
history. The framework reaches state-of-the-art performance on three
text-based recommendation tasks, namely news, quotation, and social
media post recommendation.

\subsection{Multi-modal LLM-based Sequential Recommendation Models}
\label{sec:multimodal}

A number of research works combine both textual and visual information
to develop multi-modal sequential recommendation models. \mmrec by Wu
\etal~\cite{wu2022mmrec} includes a cross-modal candidate-aware
attention network that selects relevant historical clicked news and
models user interest in candidate news. Experiments suggest that
multi-modal news information accelerates news recommendation
performance. \transrec~\cite{wang2022transrec} slightly modifies the
popular identifier-based recommendation framework by directly learning
from the raw features of the multi-modal items in an end-to-end style,
enabling transfer learning without depending on overlapped users or
items. \transrec appears to be a generic model that can be transferred
to various recommendation tasks. However, it entails a high training
cost due to its end-to-end learning method, and can only be applied to
image-only, text-only, and image-text scenarios.

To reduce \transrec's high training cost,
Fu \etal~\cite{li2023exploring} research whether an adapter-based variant
of \transrec with two task-specific neural modules inserted into each Transformer block
could perform similarly to the original fine-tuned model~\cite{wang2022transrec}.
Their findings suggest a comparable
performance between the two for both text and image recommendations.
To shed light on the modality-based versus identifier-based
recommendation architecture dilemma, Yuan \etal~\cite{yuan2023where}
conduct an empirical investigation tackling various associated
questions. They conclude that modality-based recommendation is
comparable to identifier-based, and could potentially surpass it given
the rapid NLP and computer vision advances. Still, the former is more
expensive in training, aligning with \transrec's
findings~\cite{wang2022transrec}.

\subsection{Comparison to Previous Paper and Related Work}

Our present work builds on top of our previous
paper~\cite{harte2023leveraging}, where we proposed three orthogonal
approaches of leveraging LLMs for the sequential recommendation
problem. The first one, \embeddingModel, obtains from an LLM the
embeddings of items in a session, combines them to construct a session
embedding, and maps the session embedding to the embedding space of
available catalog items in order to produce next-item recommendations.
The second approach, \promptModel, involves fine-tuning an LLM to
generate next-item recommendations. We provided session data in the
form of prompt-completion pairs, where the prompt contains all items
of each session but the last and the completion contains the last
item, that is the ground truth. At prediction time, we asked the
fine-tuned model to recommend an item for each session of the test
set. Finally, the third approach, \bertmodel, entails initializing
the embedding layer of \bertrec with LLM embeddings to enable
\bertrec to leverage the embeddings in its training process. Experimental
results on the Amazon Beauty and a Delivery Hero dataset showed that
this approach improves accuracy in terms of NDCG@20 by 15--20\%.

In the meantime, a new body of research work involving LLMs in the
sequential recommendation problem has swiftly emerged as we discuss in
Sections~\ref{sec:cross-domain-llms},~\ref{sec:domain-specific-llms},
and~\ref{sec:multimodal}. With respect to leveraging LLM embeddings
for producing recommendations, we identify no similar work to our
\embeddingModel approach. The majority of research efforts target
fine-tuning LLMs for various recommendation tasks, and result in mixed
conclusions, aligning with our \promptModel observations. Researchers
have devised a variety of methods including combining closed and
open-source LLMs~\cite{liu2023once}, using algorithms from search
engines~\cite{li2023gpt4rec}, utilizing user
profiles~\cite{gao2023chatrec}, and structuring recommendations as
instructions~\cite{bao2023tallrec}. Overall, these methods appear
mainly suitable for cold-start and data-scarce recommendation
problems, and highly depend on the applied domain and
dataset~\cite{dai2023uncovering}.

Regarding the initialization of neural sequential models with LLM
embeddings, the recent approaches presented
in~\cite{Hou2023learning},~\cite{hua2023how}, and~\cite{yuan2023where}
differ from our work in particular in terms of the goals they pursue.
\vqrec~\cite{Hou2023learning} targets cross-domain recommendations
with a novel item representation scheme, while~\cite{hua2023how}
and~\cite{yuan2023where} evaluate whether recommendation models
leveraging different modalities perform better than existing
recommendation models that rely on item identifiers. In general, there
seems to be a common conclusion that enhancing neural sequential models with LLM
embeddings advances model performance due to the enriched information
contributed by the embeddings~\cite{Hou2023learning, yuan2023where,
  zhang2023prompt}.
As we will notice, this conclusion is also supported by our own results.

In this work we propose multiple and substantial extensions to our
previous paper.
Overall, the extensions regard (i) a new breed of hybrid approaches
born from the above-mentioned approaches and from existing
recommendation models, (ii) additional LLM embedding models, (iii)
more dimensionality reduction methods, (iv) additional LLMs available
for fine-tuning and new versions of those, (v) new fine-tuning task
specifications, and (vi) more sequential recommendation models
enhanced with LLM embeddings.

\begin{figure}[t]
  \includegraphics[width=.8\textwidth]{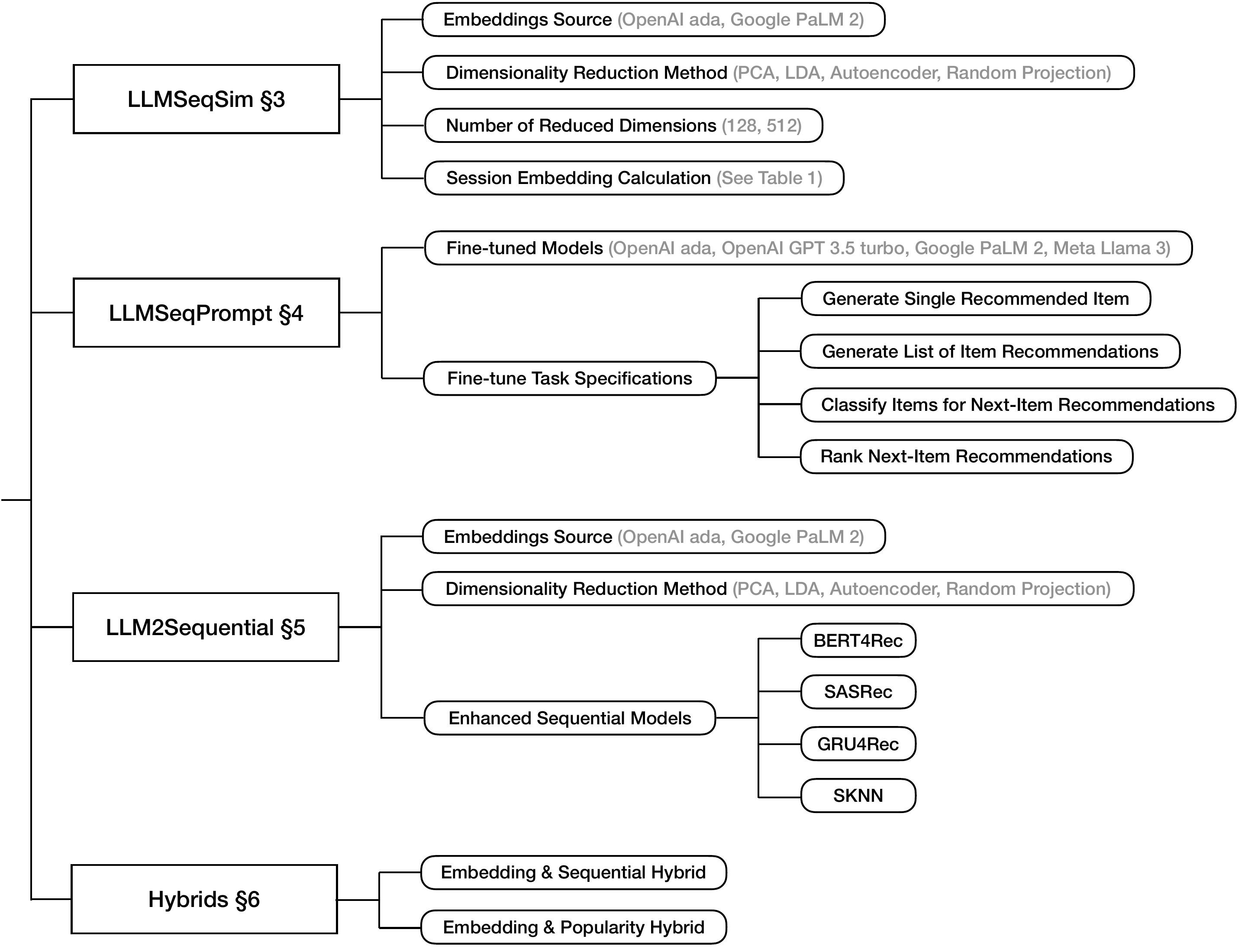}
  \caption{Layout of work. Each top-level branch of the tree shows one approach presented in
    the corresponding section. In brackets we indicate the
    alternatives we used.}
  \label{fig:work-layout}
\end{figure}

These extensions are implemented in three orthogonal plus a new breed
of hybrid approaches, shown in Figure~\ref{fig:work-layout}. Each
approach comprises a number of technical aspects for which we devise
and investigate alternative choices. We evaluate a large variety of
configurations for all approaches on the same experimentation testbed,
against the state-of-the-art sequential recommendation models, using
three datasets of different domains. In this way, we delve into each
of the fundamental approaches regarding the involvement of LLMs in the
sequential recommendation problem and assess their potential to
advance the existing state of the art and point to promising new
avenues for research. Finally, based on encouraging results, we
explore hybrid approaches by combining variations of the proposed
methods, aiming to use each individual method's dominant
characteristics in an ensemble to improve recommendations further.

\section{\embeddingModel: Semantic Item Recommendations via LLM
  Embeddings}
\label{sec:embeddings}

\begin{figure}[t]
  \centering
  \includegraphics[width=\textwidth]{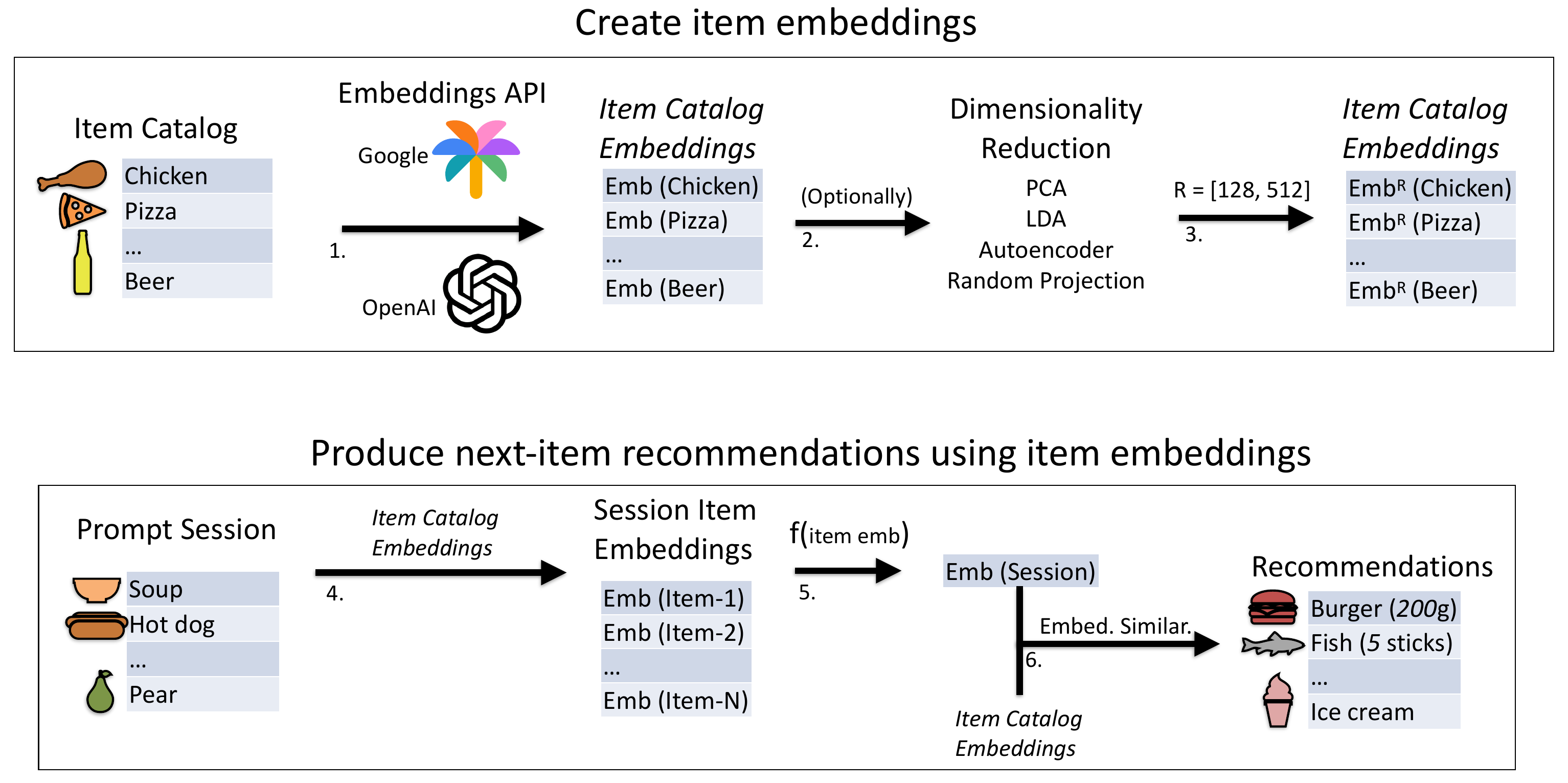}
  \caption{Semantic item recommendations via LLM embeddings.}
  \label{fig:llmseqsim}
\end{figure}

Our goal with this first approach is to explore if recommendations can
benefit from a notion of semantic similarity as provided by LLMs.
Specifically, we assume that LLM embeddings encapsulate semantic
relationships regarding items of many domains based on their massive
training data. Therefore, when we encounter a recommendation use case
for a particular domain, the LLM embeddings can position items of the
domain relative to other items, considering the items' semantics and
their relationships with relevant concepts of the domain. In this way,
embeddings can enable recommendations that stem from a deep semantic
intuition of the items' nature. In addition, putting together a
sequence of such item embeddings opens a potential for exploring a
space of recommendations that complete the sequence in novel ways not
envisioned before, as we elaborate in
Section~\ref{subsec:embedding-computation}.

In the proposed \embeddingModel (Sequential Similarity) approach, we
leverage LLM embeddings to produce recommendations in six steps. We
depict these steps in Figure~\ref{fig:llmseqsim}, applied on a food
recommendation use case. At the first step we supply metadata of the
catalog's items, such as the names of the products, to an LLM
embedding API and retrieve the corresponding embeddings. Then, at the
second step, we can optionally reduce the dimensions of the embeddings
to improve, if possible, the performance of the recommendation task.
Third, we store the embeddings of the item catalog.

When a prompt session arrives, for instance a cart of products, we
query the item catalog embeddings using the prompt's item metadata and
retrieve their embeddings at the fourth step. Then, we compute a
session embedding for each prompt session in our test set by combining
the embeddings of the individual items in the session. As a sixth
step, we compare the session embedding to the embeddings of the items
in the item catalog using cosine, Euclidean, and dot product
similarity.\footnote{The choice of the similarity measure did not
  significantly impact the results. We used cosine similarity.}
Finally, we recommend the top-$k$ products from the catalog with the
highest embedding similarity to the session embedding.

Overall, \embeddingModel shares similarities with the sequential model proposed in~\cite{li2023text}, which also relies on creating item embeddings from textual representations derived from metadata features, although not based on LLMs.
We identified four technical aspects in this processing pipeline, where
different implementation alternatives are possible. Below, in
Sections~\ref{subsec:embedding-sources}--\ref{subsec:embedding-computation}, we elaborate our investigation
of the different aspects involved in the \embeddingModel approach.

\subsection{Embedding Sources}
\label{subsec:embedding-sources}

With regard to LLM sources that provide embeddings in the first step,
we explore both OpenAI's embedding model
(\texttt{\small{text-embedding-ada-002}})\footnote{\url{https://platform.openai.com/docs/guides/embeddings/second-generation-models}}
as well as Google's PaLM2-based embedding model
(\texttt{\small{textembedding-gecko@002}}).\footnote{\url{https://cloud.google.com/vertex-ai/docs/generative-ai/embeddings/get-text-embeddings}} % embedding model based on Google PaLM2.
Our motivation for choosing these embedding models is to investigate
and uncover commonalities and differences between embeddings provided
by diverse and highly performing foundational models.

\subsection{Dimensionality Reduction Methods}
\label{sec:dim-reduction-methods}

After retrieving the embeddings, we can directly use them. However,
given that the embeddings are of multidimensional nature and
especially LLMs utilize a large number of dimensions to model
a vast number of concepts in the same space, we take the opportunity
to lower the embedding dimensions using a set of dimensionality
reduction methods that may improve the performance of the
recommendation task.

We applied four different dimensionality
reduction methods.
The first dimensionality reduction method we used is Principal
Component Analysis (PCA), a method that has proved very effective for
the initialization of the embedding layer of neural
models~\cite{harte2023leveraging}. Then, as an alternative to PCA we
also tried Linear Discriminant Analysis (LDA), which projects the data
to $k-1$ dimensions so that the projected data best fit $k$ classes.
We used the most characteristic item metadata to denote classes, such
as the name or category of each item depending on the dataset.
The third approach was to apply an autoencoder to reduce the
dimensions of embeddings, while experimenting also with $\ell2$
regularization that might further benefit the reduction by driving
minor embedding scores to zero. Finally, the fourth approach was a
random projection of embeddings~\cite{johnson1984extensions} to a
lower dimensional space with the premise that the lower space provides
a suitable approximation of the original space in terms of distances
between the points.

\subsection{Number of Reduced Dimensions}

In each of the four dimensionality reduction methods we tried
different numbers of lower embedding dimensions. The motivation behind
this exploration is to discover whether and to which extent
dimensionality reduction produces embeddings of higher quality that
lead to improvements in recommendation performance.

Our starting dimensions were 768, which is the number of dimensions of
Google embeddings, and 1536, which is the number of dimensions of
OpenAI embeddings. We tried reducing these to 128 and 512 dimensions
to balance the overall hyperparameter search space and the
performance impact of the dimensions based on observations from ad-hoc
experiments. Prior to the hyperparameter search process, we also explored 256, 64, and less than 64 dimensions manually in ad-hoc experiments, but did not identify any potential
for further investigation in low dimensions.

\subsection{Session Embedding Computation}
\label{subsec:embedding-computation}

In the third step, we compute an aggregate session embedding for each session in
our test set by combining the embeddings of the individual products in
the session. The idea is similar to that of Global Average Pooling
(GAP)~\cite{lin:2014} used in Convolutional Neural Networks (CNNs).
Before the final, top layers of a network, the inputs have been
transformed to two-dimensional structures. In CNNs, each of these
structures, called a feature map, can be understood as the result of
applying filters to the input (typically image) data. GAP takes each
feature map and produces a single value, its average, which is then
used in the following layers. In this way, a structure of
$(\mathit{height}\times\mathit{width}\times\mathit{feature\ maps})$
dimensions is transformed to a one-dimensional vector with
$\mathit{feature\ maps}$ entries. Instead of images, we deal with a
structure of $(\mathit{session\ length}\times\mathit{embeddings\
  dimension})$, which we want to reduce to a one-dimensional vector with
$\mathit{embeddings\ dimension}$ entries. This vector will pool the
session embeddings so that the resulting vector will represent the
whole session. Intuitively, as each embedding is a vector that
represents the semantics of a session item, a global average over all
item embeddings in a session will produce a vector whose semantics
will represent the average semantics of the session.

It is possible to use different pooling methods, beyond global average
pooling. We therefore try a number of different pooling strategies:
(i)~the average of the product embeddings, (ii)~a weighted average
using linear, harmonic, and exponential decay functions depending on
the position of the item in the session, and (iii) only the embedding
of the last product.\footnote{We also tried to create an aggregated
  session embedding by concatenating the plain product names and then
  querying the OpenAI embeddings API. This however led to worse
  results.} We provide the exact computation techniques of the
weighted average in Table~\ref{tab:technique-wavg}, where for each
technique we specify the decay computation at each step of a session.

\begin{table}[t]
\begin{tabular}{rl}
\hline
\textbf{Technique} & \textbf{Decay per Step} \\
\hline
Constant linear &
$1/10$
\\
Scaling linear &
$1 / |S|$
\\
Scaling quadratic &
$1 / |S|^2$
\\ [2pt]
Scaling cubic &
$1 / |S|^3$
\\
Log &
$1 / \log(i+1)$
\\
Harmonic &
$1 / i$
\\
Squared harmonic &
$1 / i^2$
\\
\hline
\end{tabular}
\caption{Computation techniques for weighted averages; $|S|$
stands for the length of session $S$.}
\label{tab:technique-wavg}
\end{table}

\section{\promptModel: Prompt-based Recommendations by a Fine-Tuned
  LLM}
\label{sub:fine-tuned-LLM}

In this approach, we inject domain knowledge to the collective
information that a base LLM incorporates. Our goal is to thereby combine 
fundamental information about the items of the domain encoded in the
LLM with domain-specific information conveyed in the fine-tuning
process.
Related approaches to fine-tune pre-trained LLMs for recommendation problems in different ways were successfully explored in \cite{bao2023tallrec,Zhang2023recommendation,wang2022transrec}.

To this end, we fine-tune an OpenAI
\texttt{\small{ada}}\footnote{\url{https://platform.openai.com/docs/guides/fine-tuning}}
model, an OpenAI \texttt{\small{GPT-3.5
turbo}}\footnote{\url{https://platform.openai.com/docs/guides/fine-tuning/what-models-can-be-fine-tuned}}
model, and the model \texttt{\small{text-bison@001}}\footnote{\url{https://cloud.google.com/vertex-ai/docs/generative-ai/models/tune-models}},
which is based on Google PaLM2. GPT succeeded \texttt{\small{ada}} as OpenAI's
recommended base model for fine-tuning, while PaLM2 is a base model
created by Google. While there is general consensus on the overall
architecture and training process of a base LLM, differences in
architectural choices, its configuration, and the training data mean
that LLMs can differ significantly.
Thus, the 
three selected models enable us to compare and contrast two fundamentally different LLMs, and to also assess the performance evolution
of the two OpenAI models. 

We fine-tune the models on training samples consisting of a prompt
(the input) and a completion (the intended output). First, we take the
training sessions and format them as prompt-completion pairs according
to the fine-tuning task specification.\footnote{We present the precise task
specifications in Sections~\ref{subsec:gen-single-item} to Section~\ref{subsec:ranking-task}.}
We also separate a small subset
of the training sessions and provide them to the fine-tuning process
as validation set. We supply 500 sessions for validation to OpenAI and
250 sessions to Google PaLM, which is the maximum number of validation
cases that the PaLM API allows. Then, we supply the prompt-completion
pairs to a model's fine-tuning API. In all configurations, we
fine-tune the model until the validation loss converges in order to
achieve the best performance.

After fine-tuning, we provide the prompts of the sessions in the test
set to the fine-tuned model and retrieve recommendations.
Finally, for generation tasks that may lead to hallucinated
recommendations, we map non-existing items back to the item catalog
through dot product embedding similarity\footnote{We use OpenAI embeddings to resolve hallucinations, which are normalized. Therefore, the dot product is the same as cosine similarity.} and obtain the final recommendations.

Besides employing a set of diverse LLMs, we propose four alternative
specifications for the fine-tuning task.
Our rationale is to examine the
behavior and performance of a model on different problem formulations,
namely \textit{generation of a single recommended
  item} (Section~\ref{subsec:gen-single-item}), \textit{generation of a list of
  item
  recommendations} (Section~\ref{subsec:generate-list-item-recommendations}),
\textit{classification}~(Section~\ref{subsec:classification-task}), and
\textit{ranking}~(Section~\ref{subsec:ranking-task}).\footnote{Budget and time limitations allowed us to apply the fine-tuning tasks of Sections~\ref{subsec:generate-list-item-recommendations},~\ref{subsec:classification-task}, and~\ref{subsec:ranking-task} to the Beauty dataset only.}
Generation, classification, and ranking
enable an investigation of fine-tuning from less to
more restricted 
settings. Fine-tuning a model on a generation task
allows the model to recommend virtually anything. In a classification
task, in contrast, a fine-tuned model is presented with a candidate
pool of item recommendations presented as categories
and selects a top-$k$
subset of those similarly to multi-class classification. Finally, the
ranking task further narrows the freedom
of the model in that the set of
available recommendations is provided in the prompt. This allows the
model to focus on the best ranking of those recommendations.

\begin{figure}[t]
  \centering
  \includegraphics[width=\textwidth]{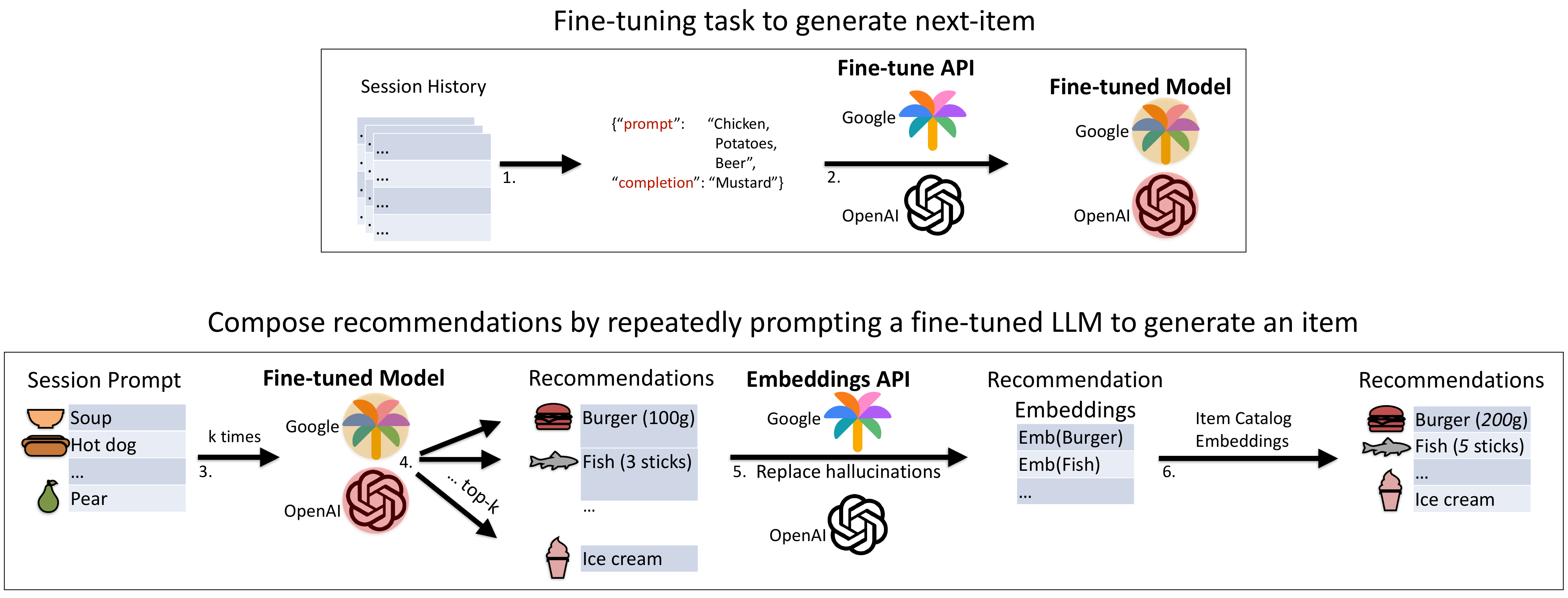}
  \caption{Next-item generation by fine-tuned LLM.}
  \label{fig:gen-next-item}
\end{figure}

\subsection{Fine-tuning Task: Generate a Single Recommended Item}
\label{subsec:gen-single-item}

In this variant, the prompt is a session, which contains a list of
item metadata, \eg, product names, except for the last item, and the
completion contains the metadata of the last item in the same session
according to the first step in Figure~\ref{fig:gen-next-item}.
At prediction time, following the fine-tuning process, we repeat a
prompt $k$ times ($k = 20)$ and collect a single recommendation
produced by the model per prompt invocation according to the fine-tune
task formulation. We use the tendency of the model to make the same
recommendation in different invocations for the same prompt as a proxy
of its confidence for the recommendation. We then collect and deduplicate the
recommendations over the $k$ invocations and rank the recommendations
by frequency of appearance. The fine-tuned LLM, being a generative
model, may also return hallucinated products. To remedy that, in the
fifth step, we retrieve the embedding of each deduplicated
recommendation 
and finally take the catalog's product that is closest in terms of
embedding similarity using the dot product.

Repeating the prompt allows us to get a number of different
recommendations from the single recommendation returned each time.
Alternatively, we can overcome the limitation of a single
recommendation per prompt, as we do next.

\begin{figure}[t]
  \centering
  \includegraphics[width=\textwidth]{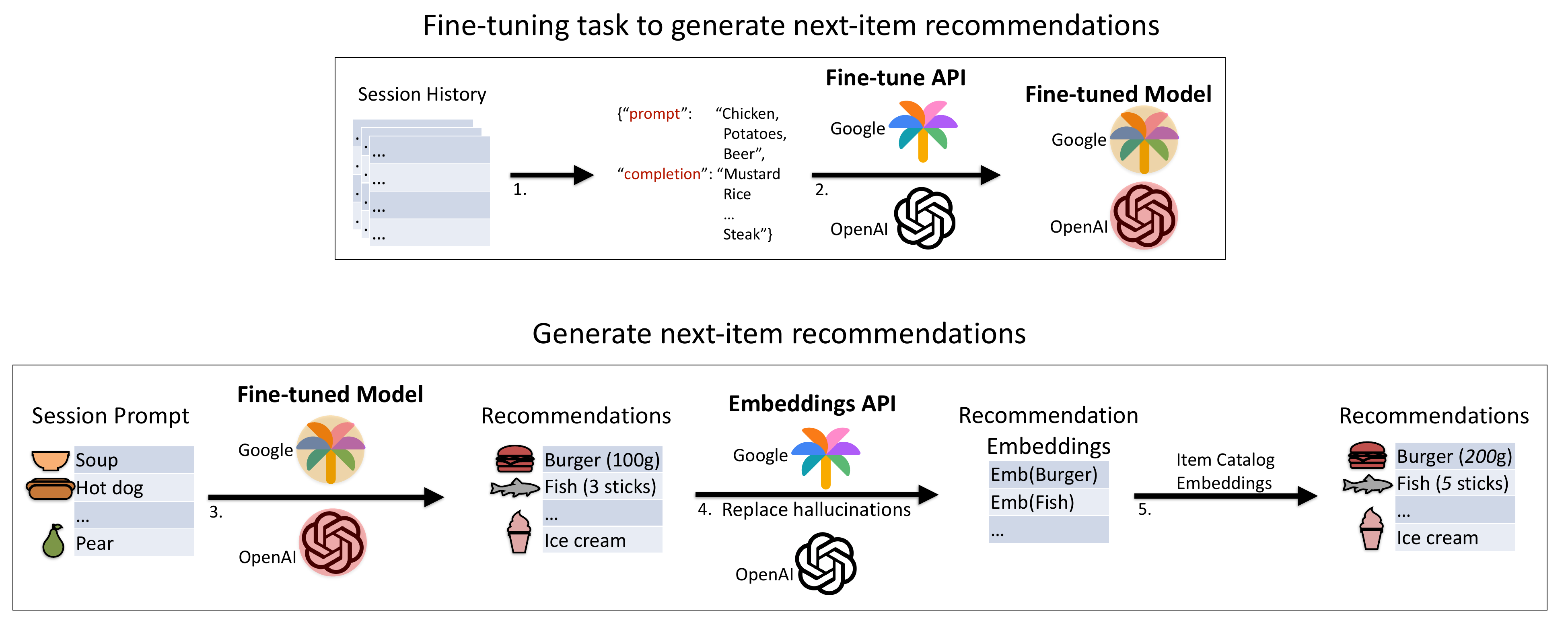}
  \caption{Recommendation list generation by fine-tuned LLM.}
  \label{fig:gen-list}
\end{figure}

\subsection{Fine-tuning Task: Generate a List of Item Recommendations}
\label{subsec:generate-list-item-recommendations}

In this approach, we teach an LLM to
embrace the role of a typical recommendation model and generate a list
of next-item recommendations ranked by order of relevance. 

To convey this task in the fine-tuning process as we depict in
Figure~\ref{fig:gen-list}, we provide in the prompt a session
containing a list of items except the last one. In the prompt
completion
we provide a list of item recommendations that is compiled by a state
of the art recommendation model. The rationale is to ask a proven
recommendation model to provide recommendations for the training
sessions and use those as prompt completions for the fine-tuning
process. In our case, we used \embeddingModel in the Beauty dataset due to its high performance.\footnote{In theory, we could also query more than one recommendation model and use the recommendations of the model that achieves the highest NDCG, or another metric.}
Finally, we insert the ground truth item at the top of the list, if not already there, and fine-tune an LLM with the same prompts and the corresponding compiled list of recommendations as completion.

At prediction time, we ask the model to produce recommendations by
passing the items of each test session except the last item. The
fine-tuned model may return hallucinated recommendations, which we map
back to the item catalog following the same method as in
Section~\ref{subsec:gen-single-item}. 

\begin{figure}[t]
  \centering
  \includegraphics[width=\textwidth]{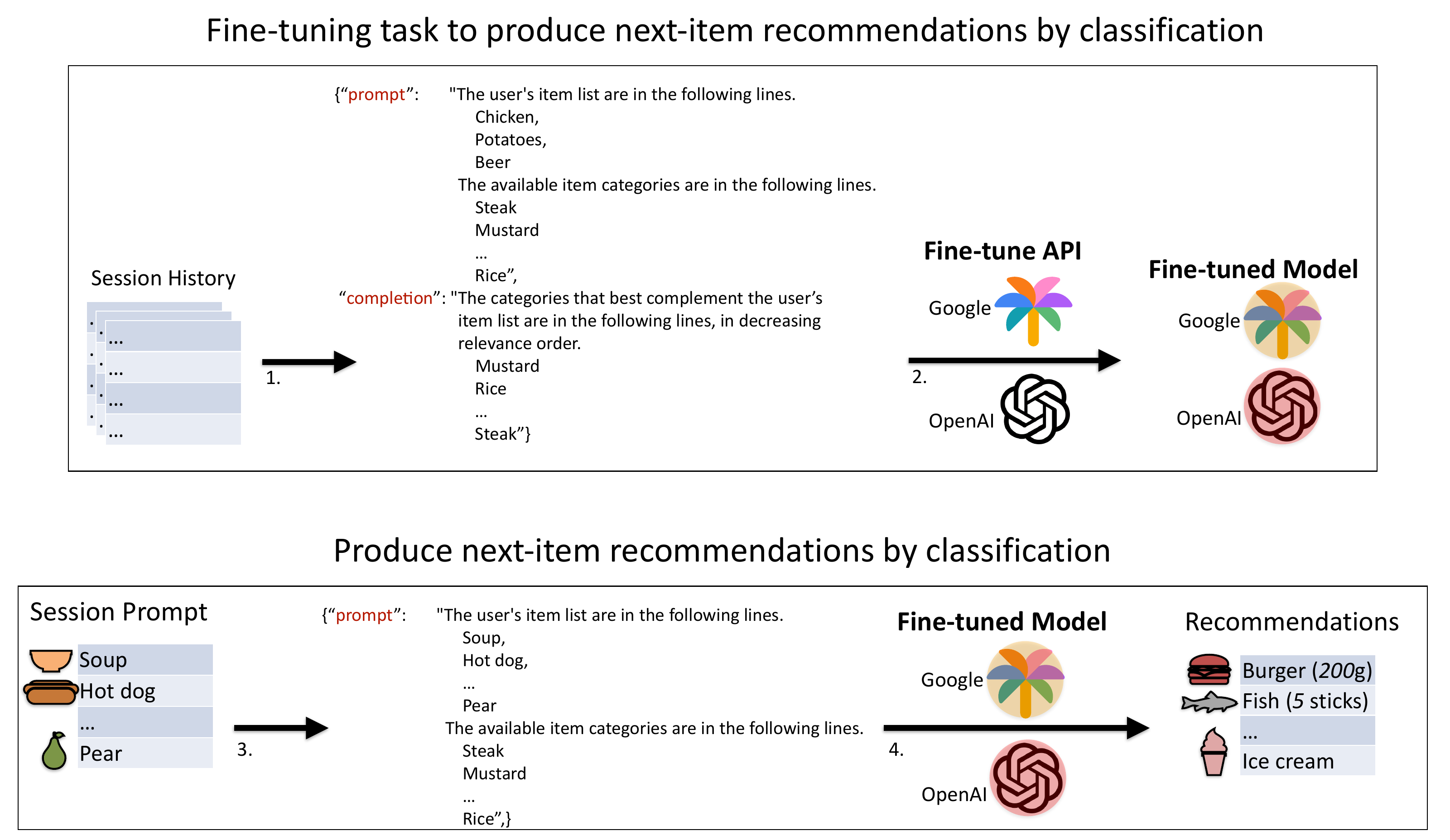}
  \caption{Multi-class classification of items by fine-tuned LLM.}
  \label{fig:classify-task}
\end{figure}

\subsection{Fine-tuning Task: Classify Items for Next Item Recommendations}
\label{subsec:classification-task}

In this approach, we explicitly provide the LLM with the eligible options for recommendation, instead of allowing it to `freely' return recommendations.
The learning goal is thus to select and rank a set of pertinent item recommendations from an overall pool of items that we pose to the model as  classification categories.
The intuition behind this formulation is that there exists a %modest
small subset of popular and diverse items in each domain that will please most customers. Restricting the set of options this way also helps us to avoid 
niche items that receive scant attention.

We state the next item recommendation problem as a multi-class classification
task as follows, see also Figure~\ref{fig:classify-task}. Prior to the fine-tuning process, we apply a
clustering algorithm, namely K-means, to group items to clusters based
on the embedding distances of their metadata.\footnote{We use product names in the Amazon Beauty dataset.}
The resulting clusters can be seen as a set of diverse item groups that are 
representative of the whole item space. We pick a relatively large
number of clusters, \ie, 200, to maximize the probability of a good
representation of semantically different items. 
Then we elevate the most popular item of a cluster to
represent the group of items in the cluster. In this fashion, we pick
an array of popular and diverse candidate items that could represent
good-quality recommendations across the total item space. Finally, for
each training session, we take the embeddings of the ground truth item
and the items that represent the classification categories and select
the top-$k$ semantically closest classification categories for each
ground truth item.

The prompts for fine tuning include the input session of items (except) the last item and the classification categories. In the corresponding completions, we include the top-$k$ classification categories that we computed in the previous step, see Figure~\ref{fig:classify-task} for an example.
At prediction time, we ask the model to classify a prompt session to top-$k$ categories, which correspond to next item recommendations.
Note that since the candidate recommendations are included in the
prompt, there is no need for an extra step to map the recommendations
back to the item space.

\begin{figure}[t]
  \centering
  \includegraphics[width=\textwidth]{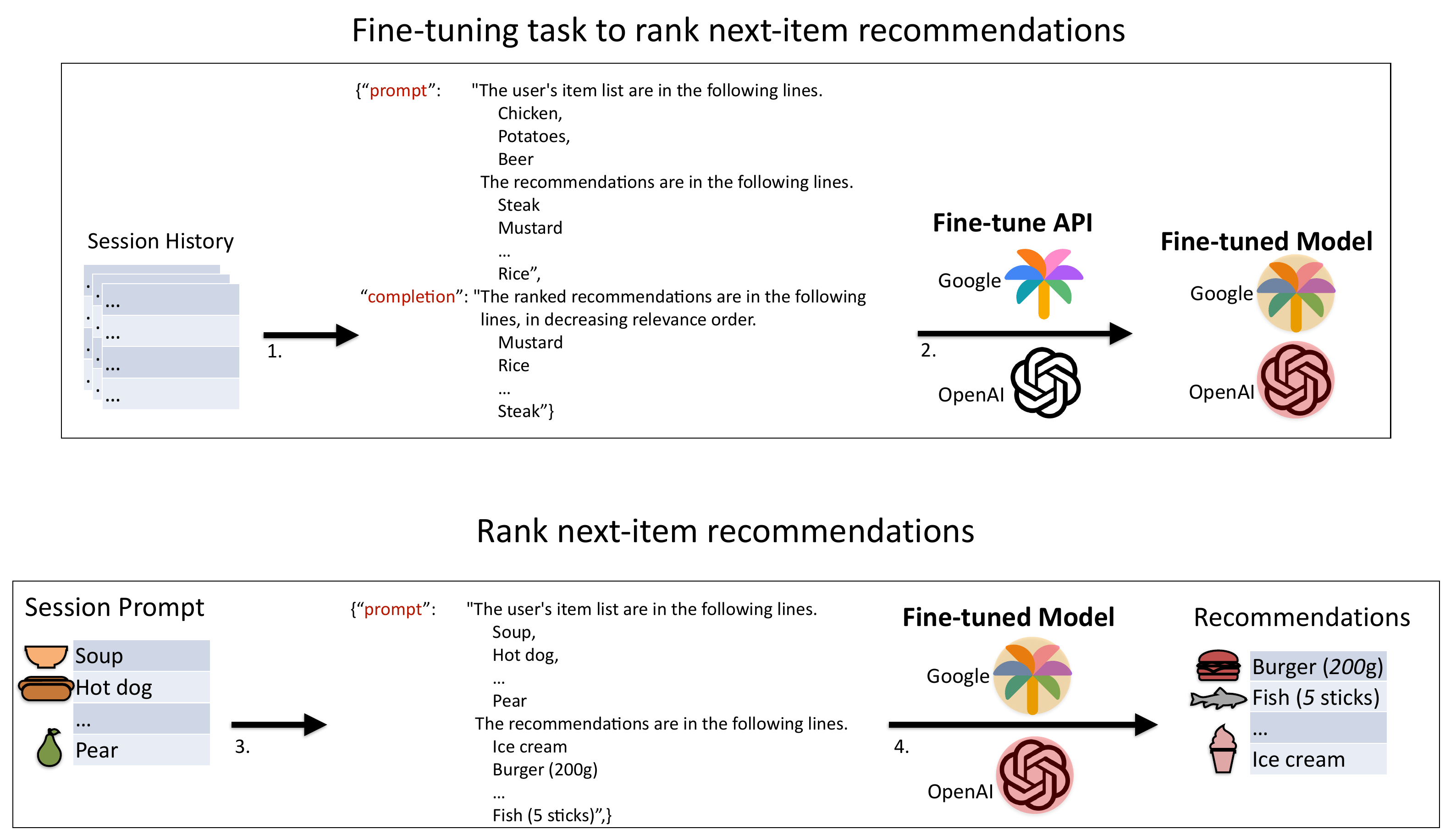}
  \caption{Rank recommendation slate by fine-tuned LLM.}
  \label{fig:rank-task}
\end{figure}

\subsection{Fine-tuning Task: Rank Next Item Recommendations}
\label{subsec:ranking-task}

In this final variant, visualized in Figure~\ref{fig:rank-task}, we
fine-tune the models to learn a ranking task. Specifically, we create a prompt that consists of \emph{(a)} a session of items except the last one, and \emph{(b)} a set of recommendations including the ground truth, that is the last item of the session.
We take the set of recommendations from an existing recommendation model depending on the dataset\footnote{We chose
  \embeddingModel for Amazon Beauty.} similarly to Section~\ref{subsec:generate-list-item-recommendations}, and shuffle
them. Then, in the prompt completion we provide the recommendations in
the order produced by the model and ensure that the ground truth item
is at the top of the list, putting or moving it there as necessary. At prediction time, we ask the fine-tuned model to re-rank the
recommendations in the prompt based on what would be the best
completion for a prompt session.

\section{\seqmodel: LLM-enhanced Sequential Models}
\label{sec:sequential}

In the third approach, our goal is to leverage the semantically-rich
item representations provided by an LLM to enhance an existing
sequential recommendation model. Existing recommendation models
typically operate on unique identifiers of the items of a domain by
learning motifs between the items from the interaction data they are
trained with. By equipping the models with meaningful item
representations we aim to enable them to learn deeper relationships
that can lead to more advanced recommendations.

We reinforce a number of sequential models with LLM capabilities,
namely \bertrec~\cite{Sun2019BERT4Rec},
\sasrec~\cite{Kang2018selfattentive}, \grurec~\cite{Hidas2016session},
and \sknn~\cite{Ludewig2018}. Of those, \bertrec, \sasrec, and \grurec
are neural recommendation models that operate on item embeddings.
Specifically, these models feature an embedding layer that
accommodates embedding representations of items. We supply item
embeddings retrieved from an LLM as input to the embedding layer of
these models. On the other hand, \sknn is a neighborhood-based
sequential model that computes session similarity based on the
presence of common items included in the sessions. In this work we
adapt SKNN to consider another notion of session similarity grounded
on the embedding similarity of items comprising a session.

Besides the consideration of different sequential models, we explore
different configurations as done for \embeddingModel described in
Section~\ref{sec:embeddings}:
\begin{itemize}
\item we use alternative LLM embedding models, \ie, those from OpenAI
  and
  Google; 
\item we explore different dimensionality reduction methods. 
\end{itemize}

Following the above-mentioned overview of our work on the enhancement
of sequential models with LLM capabilities, we organize the
description of this approach in two parts. First, we summarize each of
the sequential models we consider (Section~\ref{subsub:sequential}).
Then, we elaborate how we incorporate LLM embeddings to the workings
of the models (Section~\ref{subsub:llm-sequential}).

\begin{figure}[t]
  \centering
  \includegraphics[width=0.6\textwidth]{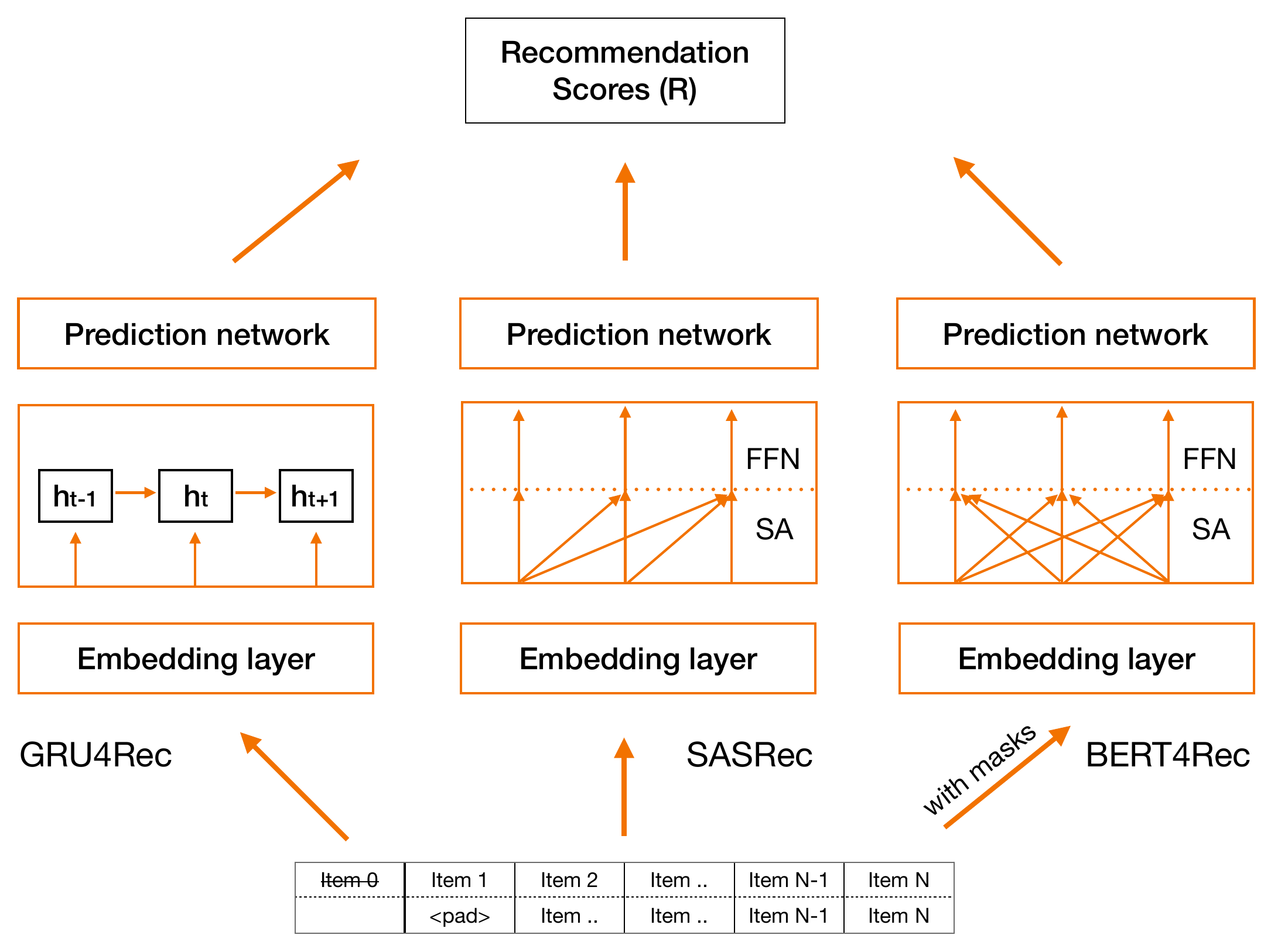}
  \caption{Architectural components of \grurec, \sasrec, and \bertrec.}
  \label{fig:gru-sas-bert-arch}
\end{figure}

\subsection{Baseline sequential models}
\label{subsub:sequential}

We provide a concise description of the baseline sequential models we
target, namely \sknn, \grurec, \sasrec, and \bertrec.
Figure~\ref{fig:gru-sas-bert-arch} depicts the main architectural
components of the neural models. We note all the details of our neural
model implementations, including any deviations from the original
model implementations, in the online
material.\footnote{\url{https://github.com/dh-r/LLM-Sequential-Recommendation/blob/main/online_material.md}}

\textbf{\sknn} (Session-based kNN) finds items to recommend by considering
sessions similar to the input session\cite{Ludewig2018}. Given a
similarity function $\mathrm{sim}(S_1, S_2)$ between two sessions
$S_1$ and $S_2$, the model considers the top-$k$ most similar
neighboring sessions $N_S$ of a session $S$ and computes the
recommendation score of an item $i$ according to the following
equation, where $\bm{1}(\cdot)$ represents the indicator function.

\begin{equation}
  \mathrm{score}(i, S) = \sum_{S' \in N_S} \mathrm{sim}(S', S) \cdot
  \bm{1}(i \in S')
\label{eq:sknn}
\end{equation}

The scores from Equation~\ref{eq:sknn} can be normalized to produce
the probability distribution $P(s_{n+1}|S)$. Though the similarity
function is intentionally left open, cosine similarity between the
sessions’ binary interaction vectors has proven to work well and lead
to competitive results in the literature
\cite{ludewiglatifiumuai2020,Kersbergen2022SerenadeScale}. % and is widely used in practice.

\textbf{\grurec} employs a recurrent neural network based on
GRU~\cite{cho2014learning}. Its overall architecture consists of an
item embedding layer, a GRU layer, and a prediction network. \grurec
is trained by sequentially processing sessions, where at each timestep
it uses the current item in the session and a hidden state computed
from previous sessions
to predict the item at the next timestep.
To elaborate, the item embedding
layer consumes the item ID $s_t$ at the current timestep $t$ and
returns its associated embedding $e_{s_t}$ from the item embedding
matrix $E$. This item embedding matrix has dimensions
$|I| \times e$, where $|I|$ is the number of items and $e$ is the
embedding dimension. Then the embedding $e_{s_t}$ is fed to a GRU
layer.
A GRU layer is stateful in that it maintains a hidden state $h_t$
while processing a session. This allows the model to use information
from the previously seen items $s_0, s_1, \ldots, s_{t-1}$ next to
item $s_t$ to predict $s_{t+1}$. Hence, the GRU layer computes $h_t$
based on its previous hidden state $h_{t-1}$ and $s_t$. Finally, the
model's output is a score $R_t$ for each item, which represents the
model’s confidence for each item that it will be the item in the
session on timestep $s_{t+1}$. Using $R_t$ we can compute next-item
recommendations by taking the items with the top-$k$ scores on the
last timestep.

In the original paper~\cite{Hidas2016session} the authors experimented
with an embedding layer but dropped it and favored a one-hot item
encoding because of slightly better performance. We kept the embedding
layer to be able to draw a meaningful comparison with other models
using LLM embeddings. We elaborate the model's implementation details
in the online material.

\textbf{\sasrec} is the first transformer
architecture~\cite{Vaswani2017attention} designed for the sequential
recommendation task. It consists of an embedding layer, one or
multiple transformer layers, and a prediction network. The transformer
layers are based on the encoder stack introduced
in~\cite{Vaswani2017attention}. In short, each of the $L$ transformer
layers consists of a self-attention (SA) and a feed-forward network
(FFN) module.
The model is trained by predicting the identity of
$s_{t+1}$ for each timestep $t$ like \grurec. However, \sasrec
computes $R$ for all timesteps $t \in [1, N]$ at once. $R$ is a
$N \times I$ matrix, where $R_{t,i}$ denotes the model's confidence
that item $i$ will be the item at position $t+1$.

Given a sequence $S$, either from the embedding layer or a preceding
transformer layer, the self-attention module takes a weighted average
of all the embeddings in the sequence, for each timestep in the
sequence. Given the output sequence of embeddings $S^L$ of the last
transformer layer, the prediction network of SASRec projects this
sequence into scores over all the items. For next-item prediction, we
therefore use $R_N$, which denotes the vector of the last item $N$ in
a session and contains the scores for all items to be the item
succeeding the last item in the session.

\textbf{\bertrec} is a state-of-the-art neural recommendation model, which
employs the transformer architecture~\cite{Vaswani2017attention} of
\bert~\cite{Devlin2019BERT}. \bert's transformer architecture
consists of an embedding layer, a stack of transformer layers, and a
prediction network. Furthermore, \bert features a masked language model
training protocol, which involves masking items at random positions
and letting the model predict their true identity.

Initially, the embedding layer embeds an input sequence of
(potentially masked) item IDs into a sequence of embeddings using both
the item ID and the item position. Then the transformer encoder layers
process the embedding sequence using a multi-head attention module and
a feed-forward network shared across all positions. Finally, the
projection head projects the embeddings at each masked position to a
probability distribution in order to obtain the true identity of the
masked item. The projection head reuses the item embeddings of the
embedding layer to reduce the model's size and to avoid overfitting.

\begin{figure}[t]
  \centering
  \includegraphics[width=\textwidth]{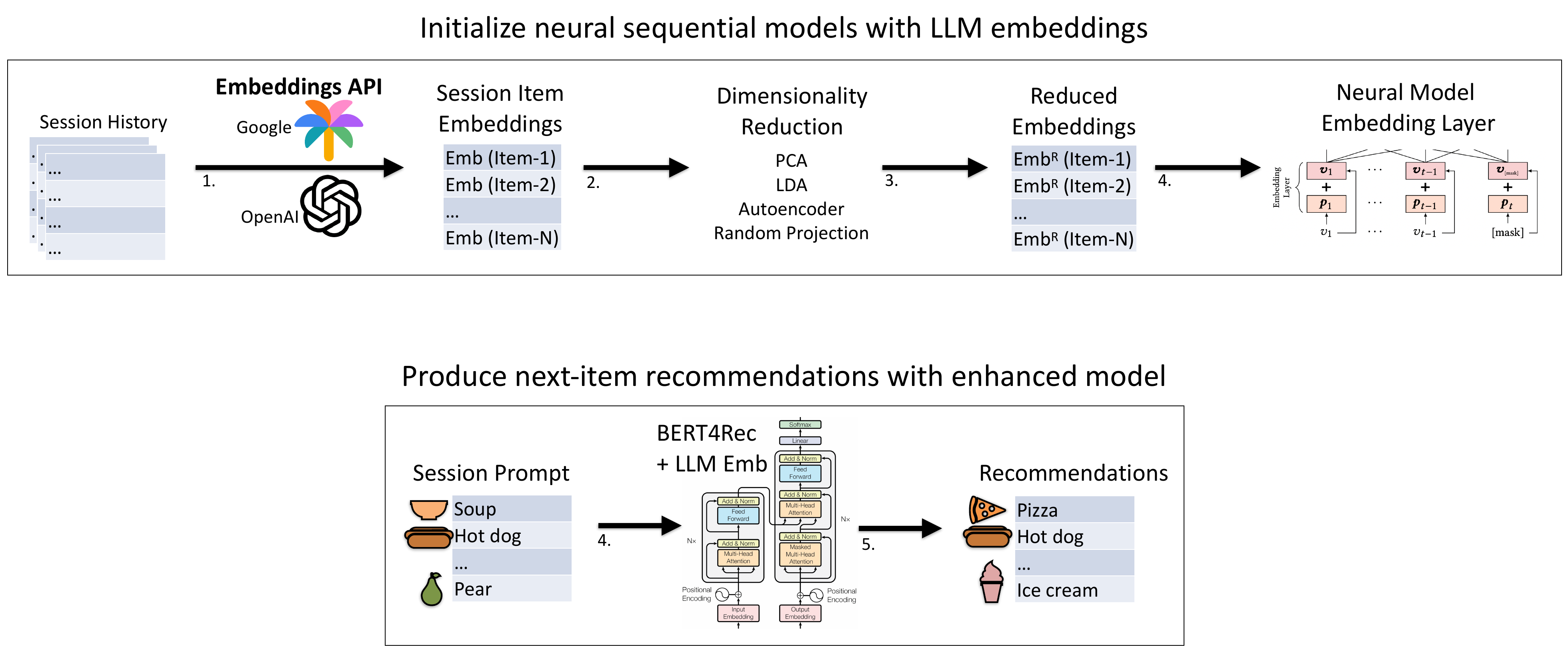}
  \caption{Neural sequential models enhanced with LLM embeddings.}
  \label{fig:llm2neural}
\end{figure}

\subsection{LLM-enhanced Sequential Models}
\label{subsub:llm-sequential}

In this approach, we incorporate LLM embeddings to two different
classes of widely applied sequential models: neural sequential models
represented by \bertrec, \sasrec, and \grurec, and
neighborhood-based ones represented by \sknn.

\paragraph{Neural Models} As we depict in Figure~\ref{fig:llm2neural}, to allow the neural
sequential models to leverage the rich information encoded in LLMs, we
employ LLM embeddings for the initialization of the neural models'
item embeddings located in the embedding layer. In order to align the
embedding dimension of the LLM embeddings (\eg, 1536) with the
configured dimension of the neural models' embedding layer (\eg, 64),
we employ and assess different dimensionality reduction methods.
Finally, we train the enhanced model in the same manner as our baseline
neural sequential model. The LLM embeddings and dimensionality
reduction methods used are described in
Section~\ref{sec:embeddings}. 

In terms of technical aspects, we note that the item embeddings in our approach are trainable and they are computed based on item metadata of the domain. For the Beauty and the Delivery Hero datasets, the metadata used to compute the embeddings are the names of the products. In the Steam dataset, the items of the domain are games. Typically, the names of the games are not closely linked with the concept of a game. Thus, we concatenate the names of the games with tags that accompany and characterize each game. The IDs of the items are not included to the input of the embedding computation.

In the overall process, the collaborative information thus enters the \seqmodel model as a sequence of IDs representing the items of the domain that appear in the sequence. In the embedding layer, the sequence of IDs becomes a sequence of embeddings in the following manner. First, the item embedding of each item in the sequence is retrieved from the embedding matrix based on the ID of the item. The item embedding is the embedding previously received by an LLM based on metadata of the item provided as input to the LLM. Second, each item embedding is summed with the positional embedding of the item’s position in the sequence to form the final embedding representation. Then, the embeddings are transformed while passing through layers of the model in the feed-forward pass where the collaborative and textual information are fused in the multilayer perceptron of the feed-forward network. Finally, through backpropagation of the gradients that minimize the cost, the weights of each layer are updated leading to the update of the embeddings in the embedding layer as well.

\paragraph{Neighborhood-based Models}
To enable neighborhood-based models, and \sknn in particular, to use
LLM embeddings, we modified the model such that it can operate on the
item embeddings. Specifically, we compute a session embedding for each
training session. To compute a score for predicting item $i$ of a
prompt session $S$ we take the session embedding $e_S$ of session $S$
and then consider the similarity of $e_S$ with the session embedding
$e_{S'}$ of each session $S'$ of the top-$k$ nearest neighboring
sessions $N_S$. We then arrive at Equation~\ref{eq:sknn-emb}, which
differs from Equation~\ref{eq:sknn} in that the latter computes the
similarity between two sessions based on the co-occurence of items in
these two sessions, while here we derive the similarity from the
session embeddings. We experiment with a number of aggregation methods
for producing a session embedding from the individual item embeddings
of a session. These aggregation methods are the same as the ones
used to produce a session embedding from individual item embeddings in
\embeddingModel (Section~\ref{sec:embeddings}). In our experiments, we
also allow the model to use one embedding combination strategy for the
training sessions and a different one for the prompt session.

\begin{equation}
  \mathrm{score}(i, S) = \sum_{S' \in N_S} \mathrm{sim}(e_{S'}, e_S) \cdot
  \bm{1}(i \in S')
\label{eq:sknn-emb}
\end{equation}

\section{Hybrids}
\label{sec:hybrids}

The three top-level approaches that we elaborated in
Sections~\ref{sec:embeddings}
(\embeddingModel),~\ref{sub:fine-tuned-LLM} (\promptModel),
and~\ref{sec:sequential} (\seqmodel) have unique
characteristics.

\begin{description}
\item[\embeddingModel] 
  relies on the semantic similarity provided by an LLM between items
  and compiles a session embedding with desirable properties across
  items of a session for producing pertinent next-item
  recommendations.

\item[\promptModel] 
  compiles fundamental and domain-specific knowledge into a single
  model potentially enabling recommendations that share information
  elements from both origins.

\item[\seqmodel] 
  is based on incorporating LLM embeddings to the sequential models.
  This approach introduces meaningful item representations to the
  training process of sequential models, aiming to boost their
  learning efficacy towards advanced recommendations.
\end{description}

In this section, we combine the unique characteristics of
\embeddingModel and \seqmodel to create stronger hybrids. The
hybrids capitalize on the notion of item popularity. Item popularity
enables segregating the item space between widely selected items, for
which a lot of information is available, and niche %obscure
items with scant
interactions. On the models' side, we have candidate models that ---as we will see---
perform very well on popular items (\eg, \seqmodel), while other
models are agnostic to item popularity and base their recommendations
on other signals (\eg, \embeddingModel). Thus, we can use the
popularity property of items to select an appropriate model based on a
cutoff popularity threshold. This threshold can be applied either
directly on the recommendations of a model
(Section~\ref{sub:hybrid-embedpopdiv}) 
or alternatively on the last item of a prompt session under the
premise that the last item is the driving factor for the next item
that will follow (Section~\ref{sub:hybrid-pop}). In this paper, we
motivate and research the potential of two concrete
hybrids\footnote{We also experimented with a hybrid based on embedding
  model variations, which did not perform better than the individual
  variations. We share the details in the online material.} based on
the aforementioned models and the notion of item popularity. We
describe these hybrids below.

\subsection{Embedding \& Sequential Hybrid}
\label{sub:hybrid-pop}

\begin{figure}[t]
  \includegraphics[width=0.95\textwidth]{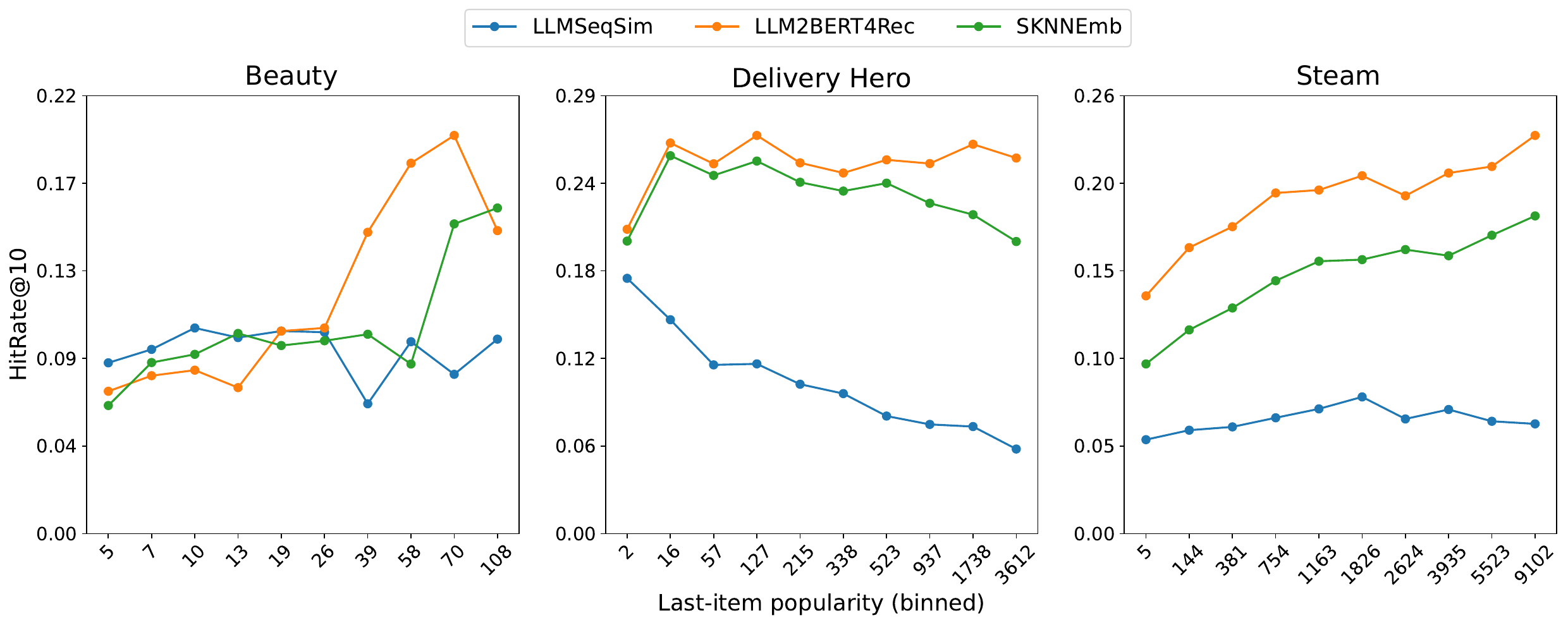}
  \caption{Performance per last item popularity.}
  \label{fig:last-item-pop}
\end{figure}

Based on the intuition that similarity-based and content-based
approaches can lead to better coverage and
novelty~\cite{Vargas:2011:RRN:2043932.2043955,JannachLercheEtAl2015,Mendoza2020Evaluating},
while many state-of-the-art recommendation models maintain a bias to
popular items, we devise a popularity-based hybrid based on two
models. The semantic item recommendation model using LLM embeddings (\embeddingModel) focuses solely at the
semantic representation of items and is oblivious to the notion of
item popularity, while \seqmodel models are generally good at
recommending popular items.
In support of this motivation, Figure~\ref{fig:last-item-pop} shows
the performance of different models with respect to item popularity.
Indeed \bertmodel and \sknnmodel exhibit higher hit rates when they
recommend for prompt sessions whose last item is popular. In contrast, the
hit rate of \embeddingModel is stable across the popularity of the
last item in the prompt sessions.

\begin{figure}[t]
  \includegraphics[width=0.95\textwidth]{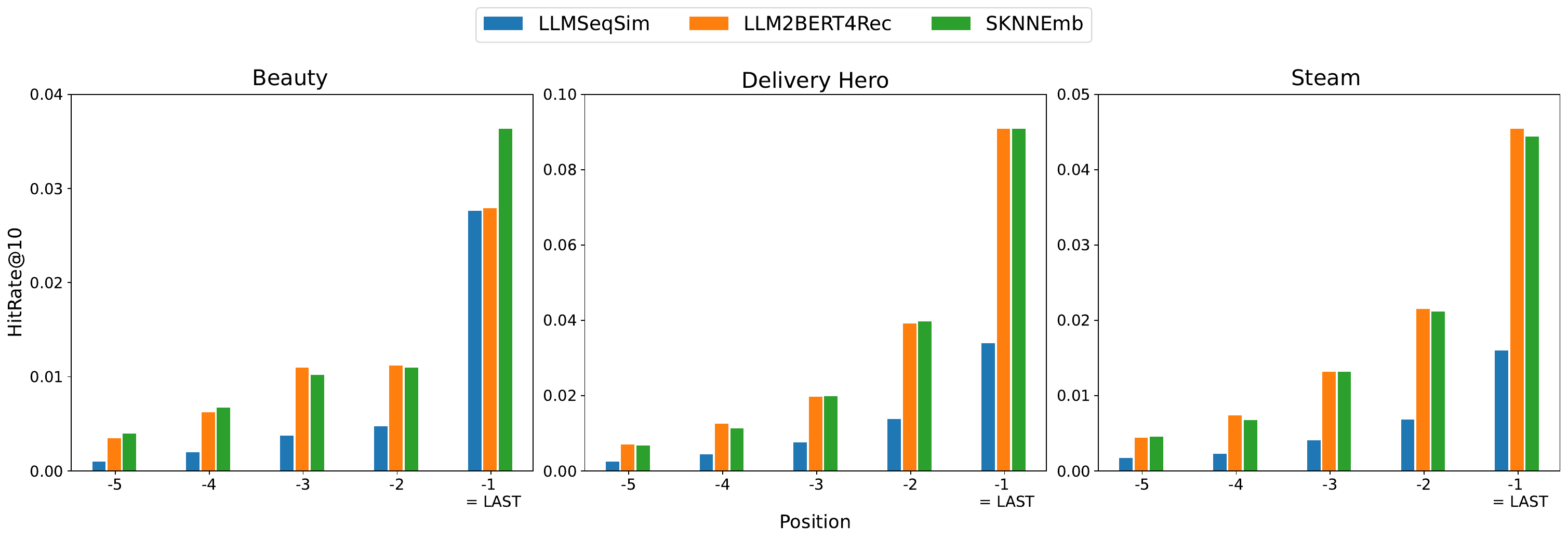}
  \caption{Performance per item position in session.}
  \label{fig:perf-item-position}
\end{figure}

In practice, we can create a hybrid where \embeddingModel is used to
recommend when encountering unpopular items and \bertmodel or
\sknnmodel are used to recommend when dealing with popular items at
the end of a prompt session. Specifically, we apply a cutoff point on
the popularity of the last item of a prompt session and use
\embeddingModel to provide recommendations for that prompt session if
the popularity of the last item falls under the cutoff point or one of
the two aforementioned sequential models otherwise. Our intuition for
using the last item as a driver for the prediction is founded on its
impact on correct recommendations according to our analysis in
Figure~\ref{fig:perf-item-position}, which shows the performance of a
model if we treat different positions in a prompt session as last. The
figure clearly shows that the item in the last position bears the
strongest impact on the performance of recommendations. We apply
hyperparameter search on the training set to identify the best cutoff
point for popularity considering the quantiles of item popularity per
dataset.

\subsection{Embedding \& Popularity Hybrid}
\label{sub:hybrid-embedpopdiv}

The LLM-embedding based recommendation model (\embeddingModel) leverages a semantically rich
representation of items, but completely lacks desirable notions of
recommendations, such as the popularity and diversity of items that
traditional recommendation models learn to embrace during their
training. Therefore, we create a hybrid based on \embeddingModel
that is also aware of popularity.

The hybrid considers popularity by enforcing a threshold on it such
that \embeddingModel does not recommend obscure items. The threshold
could be made relative to the number of options that the model has
picked for recommendations in order to include highly relevant
recommendations of low popularity.
But we did not further explore this alternative.
As with the previous hybrid (Section~\ref{sub:hybrid-pop}), we apply
hyperparameter search on the training set to determine the optimal
popularity cutoff point. We also apply a threshold on diversity in
order to enforce the semantic diversity of items in the top-$k$ list
of recommendations quantified by the items' embedding similarity. Specifically, we remove any item whose pair-wise embedding similarity with any other item in the recommendation list exceeds the threshold.
We tried both diversity alone and combined with popularity, but it did
not produce any promising results. Therefore, we proceeded with popularity only. We share all results in the online
material.

%% ===================================
\section{Experimental Evaluation}
\label{sec:experiments}
%% ===================================

In this section we describe our experimental setup
(Section~\ref{subsec:experimental-setup}), the results of our
empirical evaluation (Section~\ref{subsec:results}), and observations
from hyperparameter tuning (Section~\ref{subsec:model-tuning}). We
publicly share the code and data of our experiments to ensure
reproducibility.

\subsection{Experimental setup}
\label{subsec:experimental-setup}

The experimental setup consists of the datasets and their
pre-processing (Section~\ref{subsub:datasets}), the metrics used to
evaluate experimental results (Section~\ref{subsub:metrics}), the
models that were included in the experiments
(Section~\ref{subsub:models}), the hyperparameter tuning process for
executing models with their best configurations
(Section~\ref{subsub:hp-tuning}), the experimental setup of \srec
(Section~\ref{subsub:s3rec}), \tallrec (Section~\ref{subsub:tallrec}), and \llamaGenItemModel (Section~\ref{subsub:llama}) as well as precise information regarding
which models were executed on which datasets, including exceptions
(Section~\ref{subsub:model-executions}).

\subsubsection{Datasets and Data Splitting}
\label{subsub:datasets}

We use the public Amazon Beauty~\cite{He2016Ups}
dataset\footnote{\url{https://cseweb.ucsd.edu/~jmcauley/datasets/amazon/links.html}, Section \emph{Per-category file}.}, a novel, real-world proprietary dataset
from Delivery Hero, and the public Steam dataset\footnote{\url{https://cseweb.ucsd.edu/~jmcauley/datasets.html\#steam_data}; Version 2: Review Data and Item Metadata.} from the gaming
domain. The Beauty dataset
contains product reviews and ratings from Amazon. In line with prior
research~\cite{Anelli2022Top}, we pre-processed the dataset so that
there are at least five interactions per user and item ($\pcore=5$).
The \dhr dataset contains anonymous QCommerce sessions for dark store
and local shop orders.\footnote{QCommerce is a segment of e-Commerce
  focusing on fast delivery times on the last mile.} To better
simulate a real-world setting, we did not pre-process this dataset,
except that we removed sessions with only one interaction.
In the Steam dataset we applied $\pcore=5$ pre-processing
and removed interactions that contain items with insufficient metadata
(title, genres, and tags). Dataset statistics are provided in
Table~\ref{tab:dataset-statistics}.

\begin{figure}[t]
  \centering
  \begin{minipage}{0.48\textwidth}
    \centering
    \includegraphics[width=0.95\textwidth]{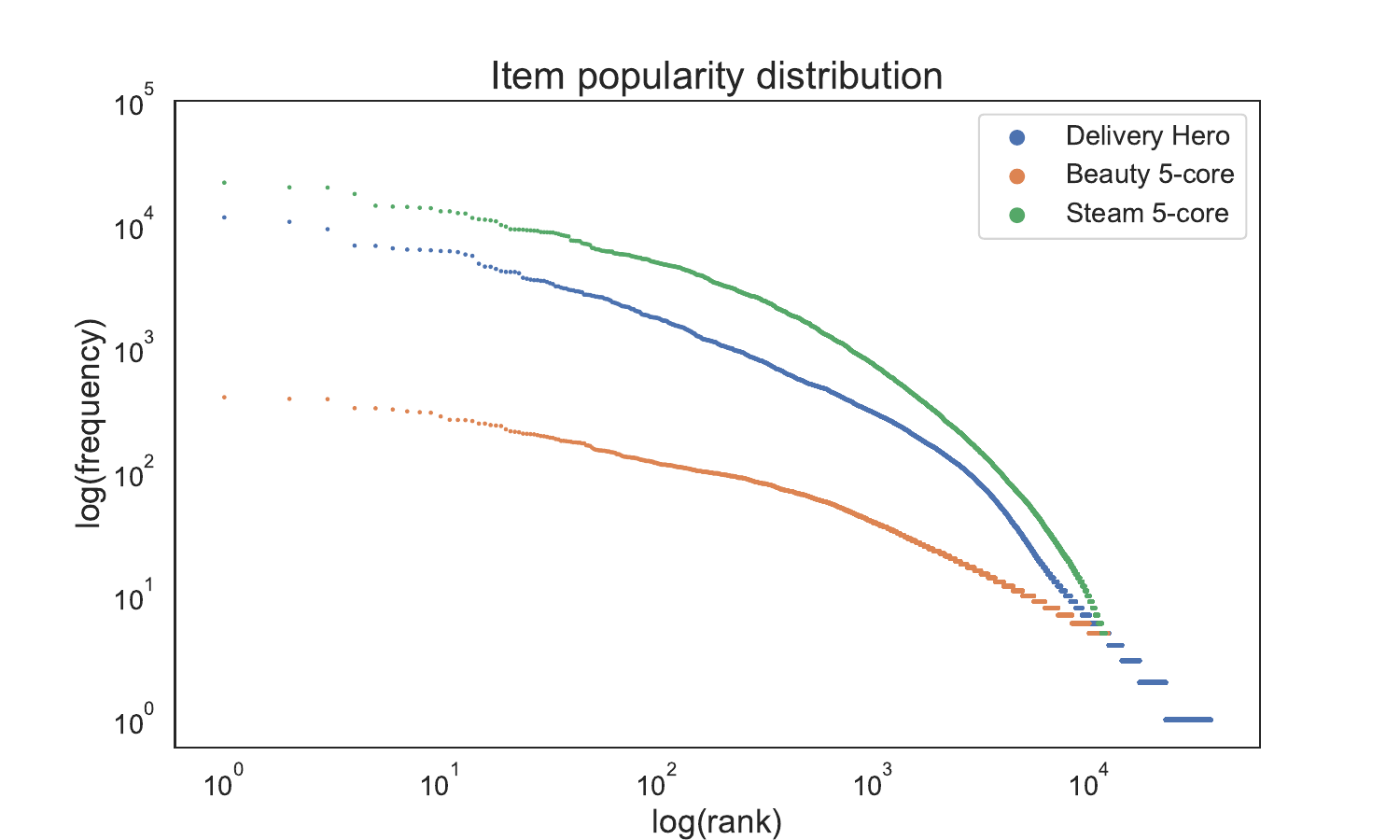}
  \end{minipage}
  \begin{minipage}{0.48\textwidth}
    \centering
    \includegraphics[width=0.95\textwidth]{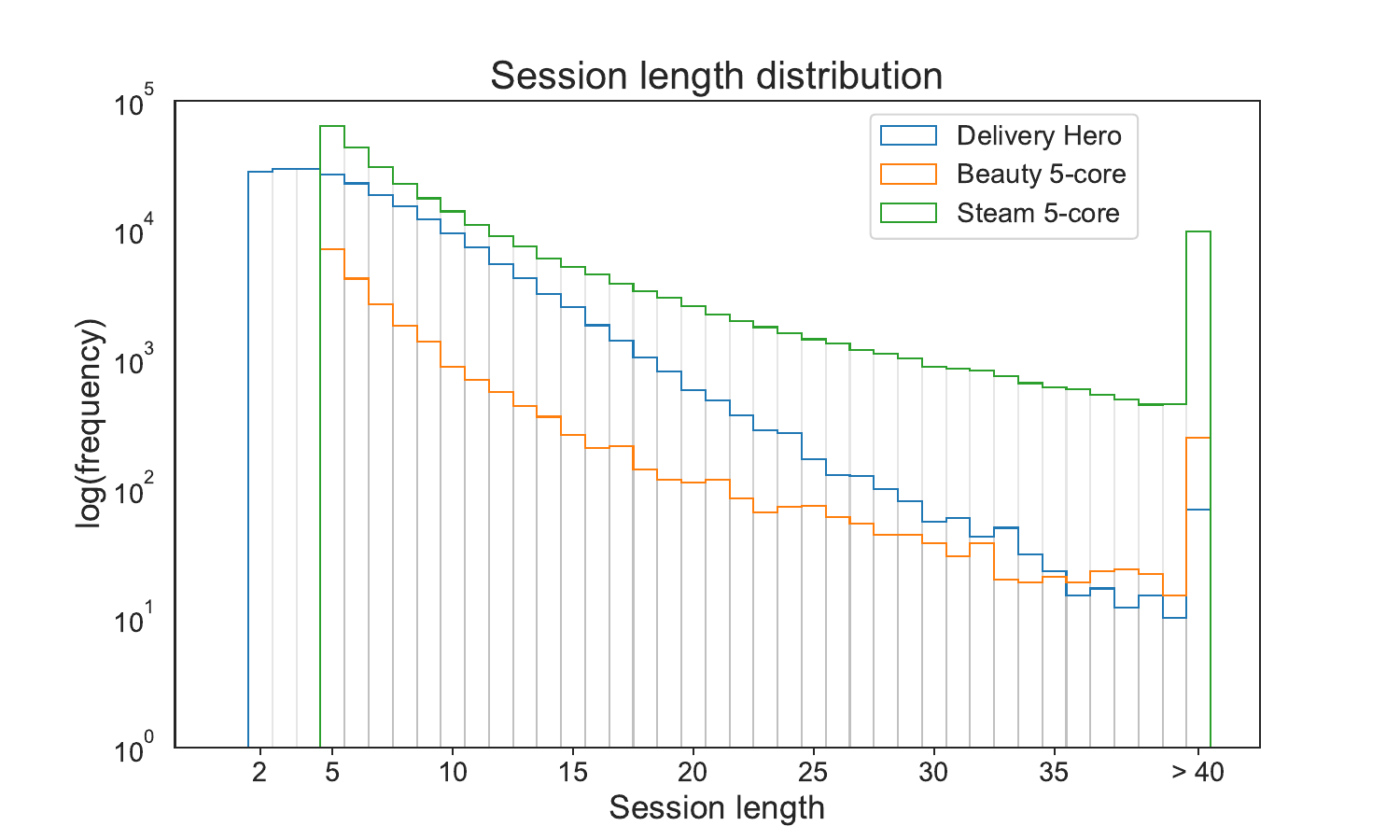} % second figure itself
  \end{minipage}
  \caption{Distribution of items ranked by popularity (left) and
    histogram of session length (right) for the datasets.}
  \label{fig:dataset-characteristics}
  \vspace{-0.5cm}
\end{figure}

We selected a diverse set of datasets to ensure that our findings are
not limited to datasets with particular characteristics. The datasets
are of different nature, originate from different domains, and have
different features.
In the left part of Figure~\ref{fig:dataset-characteristics}, we visualize
the popularity of items in the sessions. After counting the number of
occurrences of each item, we rank them in descending order (the most
popular being the first) and then plot the log of the frequency on the
log of the rank. The \dhr dataset has a tail of items that appear 1 to 4
times, as it is not $\pcore=5$. In the right part of the
figure, we show the log of the frequency of the sessions' length.
The \dhr dataset has shorter sessions than the Steam dataset, and
fewer longer sessions (those with more than 40~items) than either of
the other two datasets.

\begin{figure}[t]
  \centering
  \begin{minipage}{0.48\textwidth}
    \centering
    \includegraphics[width=0.95\textwidth]{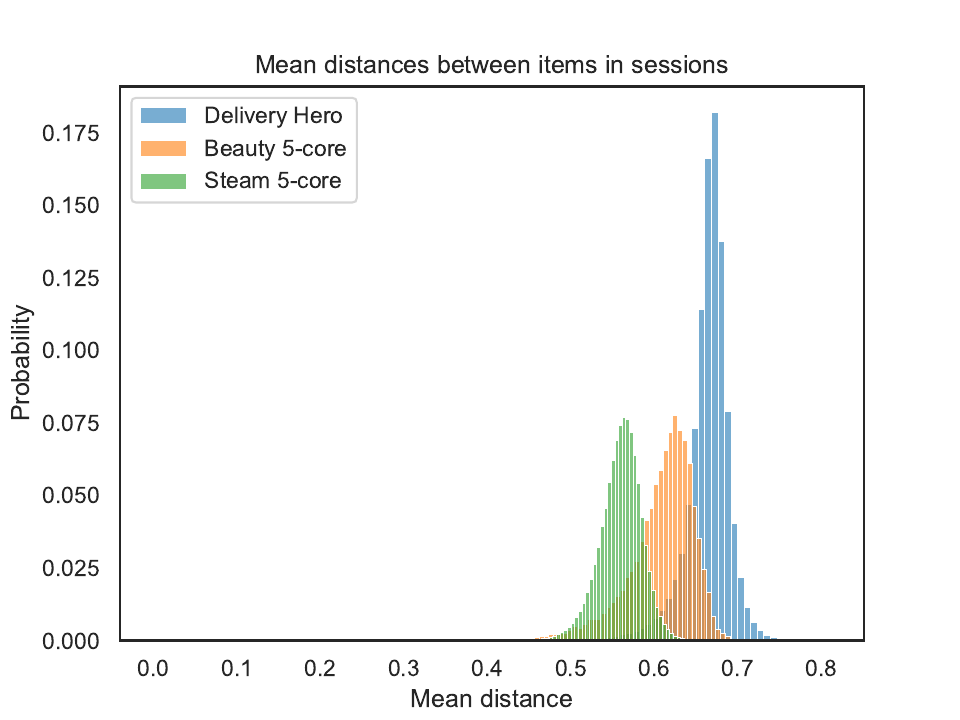}
  \end{minipage}
  \begin{minipage}{0.48\textwidth}
    \centering
    \includegraphics[width=0.95\textwidth]{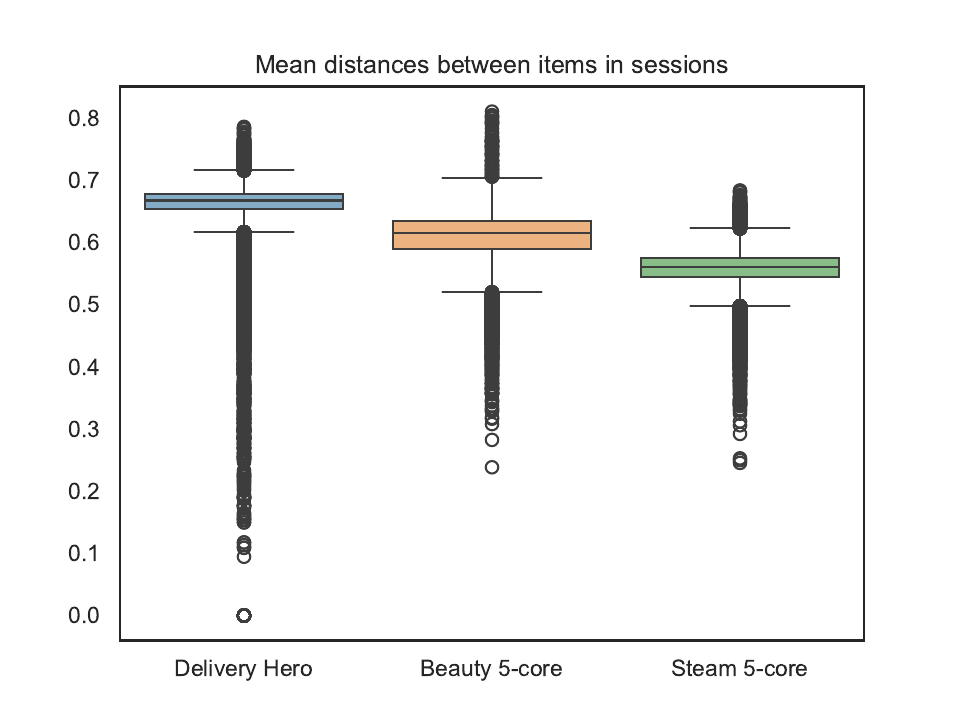}
  \end{minipage}
  \caption{Distribution of mean embedding distances for the datasets.}
  \label{fig:dataset-embedding-distances}
  \vspace{-0.5cm}
\end{figure}

Figure~\ref{fig:dataset-embedding-distances} sheds more light on the
differences between the datasets by focusing on the semantic affinity
of the items in each session (\ie, on intra-session diversity). To
find the semantic affinity of a session, we obtained the OpenAI
embeddings of its items. These being vectors, we calculated the
pairwise Euclidean distances of the embeddings of each session, and
from the pairwise Euclidean distances we then calculated their mean.
Greater distances (and their mean) correspond to smaller semantic
affinity, or more diverse session items. In this sense, the mean of
the pairwise Euclidean distances is a measure of diversity of a
session. In the histograms on the left part of
Figure~\ref{fig:dataset-embedding-distances}, we see that
the \dhr dataset has higher and less dispersed mean distances. This is
also shown in the corresponding boxplots on the right. Less diverse
sessions mean that the recommendation problem has to deal with more
similar items and thus may be easier to tackle by recommendation models.

\begin{table}[t]
\begin{tabular}{rrrrrrr}
\hline
  \textbf{Dataset} & \textbf{\# sessions} & \textbf{\# items}
  & \textbf{\# classes} & \textbf{\# interactions} & \textbf{Avg.\ length}
  & \textbf{Density} \\
\hline
Beauty $\pcore=5$ &
22,363 &
12,101 &
220 &
198,502 &
8.9 &
0.073\%
\\
\hline
\dhr &
258,710 &
38,246 &
NA &
1,474,658 &
5.7 &
0.015\%
\\
\hline
Steam $\pcore=5$ &
279,290 &
11,784 &
700 &
3,456,395 &
12.4 &
0.105\%
\\
\hline
\end{tabular}
\caption{Dataset statistics}
\label{tab:dataset-statistics}
\end{table}

To apply LDA on the data points as explained in
Section~\ref{sec:dim-reduction-methods}, it is required to provide a
set of input classes. Recall that the purpose of LDA is to project the
data associated with $k$ classes to $k-1$ dimensions so that the
projected data best fit those classes. The Beauty and
Steam\footnote{We use the genres.} datasets contain multiple
categorical keywords per item. We sort these keywords and
concatenate them using underscore as a separator to obtain unique classes
per item. The Steam dataset also contains tags, which are provided by
users. We process these in the same manner as the genre keywords. The
\dhr dataset does not readily include metadata that we can use as
classes, so we do not apply LDA to it.

To create a training and test set in a sound way, we first split a
dataset containing sessions temporally such that all test sessions
succeed train sessions in time. Then in the test set, we adopt the
leave-one-out approach as in~\cite{Kang2018selfattentive,
  Sun2019BERT4Rec}, where all but the last interaction of each session
represent the prompt (\ie, the ongoing session), and the last
interaction serves as the ground truth.

\subsubsection{Metrics}
\label{subsub:metrics}

We use the standard ranking accuracy metrics NDCG, MRR, and HitRate at
the usual cutoff lengths of 10 and 20.
Furthermore, we consider the following \emph{beyond-accuracy} metrics
to obtain a more comprehensive picture of the performance of the
different algorithms: catalog coverage, serendipity, and novelty.
\emph{Catalog coverage} represents the fraction of catalog items that
appeared in at least one top-$n$ recommendation list of the users in
the test set \cite{JannachLercheEtAl2015}. \emph{Serendipity} measures
the ratio of correct recommendations per user that are %
not recommended by a popularity baseline~\cite{Ge2010beyond}.
\emph{Novelty} computes the negative log of the relative item
popularity, or self-information~\cite{Zhou2010solving}.

\subsubsection{Models}
\label{subsub:models}

We include both session-based algorithms of different families,
\grurec~\cite{Hidas2016session}, and \sknn~\cite{Jannach2017when}, as
well as three state-of-the-art sequential models,
\bertrec~\cite{Sun2019BERT4Rec},
\sasrec~\cite{Kang2018selfattentive}, and \srec~\cite{CIKM2020-S3Rec}.
Notably, \srec introduces a custom pre-training and fine-tuning method that allows a meaningful comparison to our proposed LLM-based approaches, especially with respect to the LLM embedding initialization of neural models (Section~\ref{sec:sequential}) and finetuning of LLMs (Section~\ref{sub:fine-tuned-LLM}).
We also provide a comparison with a fine-tuned open-source Llama~\cite{touvron2023llama} LLM.
In addition, we evaluate \tallrec~\cite{bao2023tallrec}, a state of the art LLM-based approach for sequential recommendations that structures the recommendations as instructions and
tunes an LLM through an instruction-tuning process. Both Llama and \tallrec are closely related to our proposed LLM fine-tuning approaches (Section~\ref{sub:fine-tuned-LLM}).
Additionally, we initialize neural models with embeddings retrieved from a pre-trained BERT~\cite{Devlin2019BERT} model to compare against our proposed initialization approach using embeddings retrieved from OpenAI GPT and Google PaLM (Section~\ref{sec:sequential}).
We refer to those models as \bertbertrec, \bertsasrec, and \bertgrurec.
Furthermore, we tested all variants of the \sknn nearest-neighbor method proposed in~\cite{Ludewig2018} and
report the results of the best SKNN variant in this paper. The complete results can be found in the online  material.
Moreover, we include the four LLM-based approaches proposed in Sections~\ref{sec:embeddings},~\ref{sub:fine-tuned-LLM}, ~\ref{sec:sequential}, and~\ref{sec:hybrids}.
Finally, we include a popularity-based baseline (MostPopular) in the
experiments.\footnote{The consideration of additional traditional baselines and most recently published sequential models is left for future work.}

\begin{table}[t]
\centering
\resizebox{\textwidth}{!}{
\begin{tabular}{llrrrl}
\hline
\textbf{Model} &
\textbf{Sec.} &
\multicolumn{3}{c}{\textbf{\# Trials}} &
\textbf{Description}
\\
&
&
\textbf{BT} &
\textbf{DH} &
\textbf{ST} &
\\
\hline
\embeddingModel &
\ref{sec:embeddings} &
591 & 620 & 631 &
Semantic item recommendation model based on OpenAI or Google embeddings
\\
\promptGenItemModel &
\ref{subsec:gen-single-item} &
16 & 14 & 4 &
Fine-tuned model to generate the next item
\\
\promptGenListModel &
\ref{subsec:generate-list-item-recommendations} &
5 & -  &  - &
Fine-tuned model to generate next-item recommendations
\\
\promptClassModel &
\ref{subsec:classification-task} &
5 & - &  - &
Fine-tuned model to classify items for next-item recommendations
\\
\promptRankModel &
\ref{subsec:ranking-task} &
5 & - & -  &
Fine-tuned model to rank next-item recommendations
\\
\bertrec &
\ref{subsub:sequential} &
661 & 52 & 57 &
Bidirectional neural model based on the transformer architecture
\\
\sasrec &
\ref{subsub:sequential} &
413 & 167 & 146 &
Unidirectional neural model based on the transformer architecture
\\
\grurec &
\ref{subsub:sequential} &
622 & 168 & 164 &
Recurrent neural model based on Gated Recurrent Unit
\\
\sknn &
\ref{subsub:sequential} &
553 & 377 & 291 &
Neighborhood-based sequential model
\\
\bertmodel &
\ref{subsub:llm-sequential} &
1074 & 118 & 95 &
\bertrec initialized with LLM embeddings
\\
\sasrecmodel &
\ref{subsub:llm-sequential} &
1280 & 272 & 226 &
\sasrec initialized with LLM embeddings
\\
\grurecmodel &
\ref{subsub:llm-sequential} &
1001 & 381 & 256 &
\grurec initialized with LLM embeddings
\\
\sknnmodel &
\ref{subsub:llm-sequential} &
767 & 218 & 128 &
\sknn utilizing LLM embeddings to compute session similarity
\\
\popHybrid &
\ref{sub:hybrid-pop} &
72 & 72 & 72 &
Hybrid of two models that perform well on popular \& unpopular items respectively
\\
\embeddingPop &
\ref{sub:hybrid-embedpopdiv} &
60 & 60 & 60 &
Semantic item recommendation model enriched with popularity
\\
\hline
\end{tabular}
}
\caption{The recommendation models used in the experiments including
  references to the section where they are described, the number of
  trials carried out for each dataset (BeauTy, Delivery Hero, STeam)
  per model in the hyperparameter tuning process, and a short
  description of each model.}
\label{tab:experiments-map}
\end{table}

\subsubsection{Hyperparameter Tuning}
\label{subsub:hp-tuning}

We systematically tuned all models (except \promptModel, \llamaGenItemModel, \srec, and \tallrec, which we discuss separately) on three
validation folds with the Tree Parzen Estimator (TPE)
sampler~\cite{Bergstra2011Algorithms}, and the total NDCG@20 across
the folds as the optimization goal. We chose to optimize for 20
recommendations to provide enough room for the models to manifest
their performance differences, although top-10 and top-20 metrics are
usually closely related.
In the beginning, we explore 40
random configurations, which helps avoid local minima, and then
continue by evaluating the suggestions by the TPE sampler. We let the
hyperparameter search run for 72 hours, but started multiple hyperparameter search experiments in parallel for each model (two for vanilla neural models and four for \seqmodel models) in order to minimize the duration of experiments.
We also allow for early
stopping of the hyperparameter search if the optimization objective
has not been improved for 100 trials. Finally, we prune a trial if its objective value (\ie, NDCG@20)
belongs to the bottom 20\% of all tested trials up to that point,
allowing us to accelerate the hyperparameter search process. We allow
pruning only after ten different configurations of a model have been
tried.

Overall, our chosen hyperparameter tuning period with early stopping and pruning enabled allows a wide search of the hyperparameter search space for the various models. Table~\ref{tab:experiments-map}
contains the number of different configurations tried per model.
Note that the \seqmodel neural models manifest a considerably higher number of trials than the respective vanilla neural models in order to account for the two additional hyperparameters they include, i.e. the embedding models and the dimensionality reduction methods, which result in a search space that is eight times larger.

The folds themselves are created by splitting the training data into
validation-training and validation-test sets. We do this in a
temporal fashion (same as for the train-test split) by splitting the
training sessions into four bins, where each bin contains 25\% of the
training sessions in chronological order. The first fold uses the
first bin as the validation-training data, and the second bin as
the validation-test data. Then the second fold uses the first two
bins as the validation-training data, and the third bin as the
validation-test data. Similarly, the third fold uses the first three bins as
the validation-training data, and the fourth bin as the validation-test
data. By using less data in the first few folds (in comparison to
cross-validation), we can more quickly evaluate a configuration as well as
preserve the temporal split in our hyperparameter search process.

For the fine-tuning step of \promptModel variants we stick to the
default parameters, such as learning rate multiplier and number of
epochs, recommended by the LLM providers. In the prediction step where
we query a fine-tuned model to provide recommendations, we vary the
temperature of the model. This hyperparameter, which in GPT takes
values between~0 and~2, controls the randomness of the model's output.
With a lower temperature the model produces more %mode
deterministic
completions, while higher temperature makes the completions more
random. Note that we treat the different LLMs used in the fine-tuning,
\ie, OpenAI ada, OpenAI GPT, and Google PaLM, as hyperparameters and
report the best performance per task variant. Thus, the number of trials for an \promptModel variant in Table~\ref{tab:experiments-map} represents the temperature configurations tried for all LLMs. The examined
hyperparameter ranges and optimal values for each dataset are reported
in the online material. 

Finally, the hyperparameter tuning of the hybrids involves the following main parameters:
embedding source, 
embedding dimensionality reduction, 
sequential model 
(for the \popHybrid hybrid),
and popularity threshold value. 

\subsubsection{Training and Evaluation of \srec}
\label{subsub:s3rec}
\color{black}
The \srec model is different in nature from other models, because it is a pre-training based approach. Nonetheless, we were interested how this particular model would fare in a comparison with the other models in our study. In terms of datasets, we included experiments only for the Beauty and Steam datasets, as these  include metadata about item attributes that are required to pre-train \srec. Using the original code of \srec provided by the authors, we pre-trained \srec for 10 epochs, fine-tuned it for 200 epochs with early stopping enabled. As for the other models in our comparison, we used the NDCG and the hit rate as evaluation measures. We note that we determined the metric values using the evaluation protocol from the original paper. \srec's evaluation protocol involves using each session's second-to-last item for validation during fine-tuning. The last item of each session is used for testing. We note that the evaluation protocol (and code) used for the other models, which is more common in the literature, is slightly different and more challenging for the algorithms, as we hide entire sessions from the validation and training data. As a result, the results reported for the \srec model are expected to be higher than what we would observe if the exact same protocol would be applied. As our results will show, however, the accuracy results obtained with \srec fall behind the other models, even though more information was available than for the other models.

\subsubsection{Training and Evaluation of \tallrec}
\label{subsub:tallrec}
For \tallrec, we present experiments only for the Beauty dataset because the model would require many weeks of execution on the  Delivery Hero and Steam datasets.
Regarding the model's training, we used the same preprocessing and train-validation-test split for Beauty as for \promptGenItemModel (Section 4.1).
To make \tallrec comparable to our work, we converted TALLRec's binary classification problem into a recommendation generation problem.
We also used the same system prompt, user prompt template, and assistant prompt template as in \promptGenItemModel but we adapted it to TALLRec's format.\footnote{\url{https://github.com/SAI990323/TALLRec/blob/main/finetune_rec.py\#L301}}
We then trained \tallrec on an EC2 g5.12xlarge instance with 4x24GB A10G GPUs using the default hyperparameter setup provided by the authors.\footnote{\url{https://github.com/SAI990323/TALLRec/blob/main/shell/instruct_7B.sh}}

For the evaluation, we used the same protocol as for \promptGenItemModel. After resolving out-of-memory issues, we asked the model to produce top-20 recommendations for each test case by repeating each test prompt 20 times. Then, we mapped each recommendation of the model that is not in the product catalog to the closest catalog product via OpenAI embeddings. Finally, we computed the evaluation metrics that we report in the paper.

\subsubsection{Fine-tuning and Evaluation of \llamaGenItemModel}
\label{subsub:llama}
We fine-tuned \llamaGenItemModel, specifically \texttt{\small{meta-textgeneration-llama-3-8b}}, %on AWS
on the Beauty and Steam datasets using the LoRA approach for five epochs with 0.0001 learning rate, 32 alpha, 0.05 dropout, and we set the \emph{r} parameter to 8. In addition, we used 20\% of the dataset as validation set. Resource constraints did not allow us to conduct experiments on the Delivery Hero dataset. On the same grounds, we fixed the temperature of the model to 1 and the \emph{top\_p} parameter to 0.25, which perform best in practice according to our experience.

\color{black}

\subsubsection{Model Executions on Datasets}
\label{subsub:model-executions}

In principle, we ran all models on all datasets except for fine-tuning
models due to budget and time limitations.
Specifically, we applied \promptGenListModel, \promptClassModel,
\promptRankModel, and \tallrec only on the Beauty dataset, and \srec and \llamaGenItemModel only on the Beauty and Steam datasets. In addition, we only
fine-tuned OpenAI \texttt{\small{ada}} on \promptGenItemModel because it is no
longer available for fine-tuning.

Besides fine-tuning experiments,
we executed again all experiments of the previous paper~\cite{harte2023leveraging}
and in fact managed to tune better a number of baseline models on the
\dhr dataset, of which \bertrec and \sknn exhibited high performance gains.
The performance of \sasrec and \grurec also improved moderately.
That said, we did observe a small decrease to the performance of \sknn and \bertrec
in the Amazon Beauty dataset, while \grurec's performance saw a small increase.
We attribute the slightly lower performance to a potential modest instability caused
by the size of the dataset and to the stochasticity of the
hyperparameter search process.

\subsection{Results and Discussion}
\label{subsec:results}

Tables~\ref{tab:results-beauty},~\ref{tab:results-DH}, and~\ref{tab:results-steam} depict the results obtained for the Amazon Beauty, \dhr, and Steam datasets on the hidden test set, respectively. The rows in each table are sorted in descending order according to the NDCG values at 20.
We discuss the experimental results in terms of accuracy and beyond accuracy metrics in Sections~\ref{subsub:accuracy} and~\ref{subsub:beyond-accuracy}, respectively.

\subsubsection{Accuracy metrics}
\label{subsub:accuracy}

The key observation of the experimental results is that \emph{LLM
  embeddings boost recommendation model performance in terms of
  accuracy metrics across models and datasets}. Specifically,
sequential models enhanced with LLM embeddings score the best
performance across all three datasets demonstrating the superiority of
LLM embeddings for the models' learning and training routines. In
particular, \sasrecmodel and \bertmodel maintain consistently the
best performance and alternate on the top spots marking around 45\%
average improvements in NDCG@20 over their vanilla counterparts on
Beauty, and 9\% on the \dhr dataset.
On Steam, the performance boost is steady but
marginal across all sequential models. One potential reason for that
could be the quality of embeddings for this particular dataset. As a
gaming dataset, Steam features game titles that are semantically
non-discriminative. Although we enriched the title with tags that
accompany the games, such as \texttt{\small{action game}} and
\texttt{\small{multi-player}}, these are commonly shared across the
games and, therefore, their contribution is ambivalent. Finally,
\grurecmodel achieves a 20\% increase in NDCG@20 on Beauty over
\grurec, but only marginal improvements in the other two datasets.

\begin{table}[t]
\resizebox{\textwidth}{!}{
\begin{tabular}{lrrrrrr|rrrrrr}
\hline
\multicolumn{1}{c}{} &
\multicolumn{6}{c}{\textbf{Top@10}} &
\multicolumn{6}{c}{\textbf{Top@20}}                                                                   \\
\multicolumn{1}{c}{\multirow{-2}{*}{\textbf{Model}}} &
\textbf{nDCG} &
\textbf{HR} &
\textbf{MRR} &
\textbf{CatCov} &
\textbf{Seren} &
\textbf{Novel} &
\underline{\textbf{nDCG}} &
\textbf{HR} &
\textbf{MRR} &
\textbf{CatCov} &
\textbf{Seren} &
\textbf{Novel}
\\
\hline
%\rowcolor{gray!20}
\sasrecmodel &
    \textbf{0.045} &
    \textbf{0.083} &
    0.034 &
    0.204 &
    \textbf{0.081} &
    11.660 &
    \textbf{0.054} &
    \textbf{0.118} &
    0.036 &
    0.300 &
    \textbf{0.112} &
    11.844
\\
\bertmodel &
    0.042 &
    0.080 &
    0.031 &
    0.226 &
    0.076 &
    11.645 &
    0.052 &
    \textbf{0.118} &
    0.034 &
    0.328 &
    0.111 &
    11.823
\\
\popHybrid &
    0.040 &
    0.074 &
    0.029 &
    0.349 &
    0.072 &
    11.729 &
    0.050 &
    0.116 &
    0.032 &
    0.514 &
    0.110 &
    11.892
\\
\embeddingModel &
    0.043 &
    0.068 &
    \textbf{0.036} &
    \textbf{0.761} &
    0.068 &
    \textbf{13.755} &
    0.049 &
    0.090 &
    \textbf{0.037} &
    \textbf{0.889} &
    0.090 &
    13.796
\\
\embeddingPop &
    0.042 &
    0.065 &
    0.035 &
    0.464 &
    0.065 &
    13.067 &
    0.049 &
    0.093 &
    0.037 &
    0.496 &
    0.092 &
    13.098
\\
\sknnmodel &
    0.039 &
    0.070 &
    0.030 &
    0.476 &
    0.068 &
    11.855 &
    0.047 &
    0.101 &
    0.032 &
    0.689 &
    0.096 &
    12.090
\\
\vsknn &
    0.036 &
    0.063 &
    0.028 &
    0.399 &
    0.059 &
    11.211 &
    0.043 &
    0.090 &
    0.030 &
    0.613 &
    0.083 &
    11.481
\\
\promptRankModel &
    0.037 &
    0.068 &
    0.027 &
    0.744 &
    0.068 &
    13.727 &
    0.042 &
    0.089 &
    0.028 &
    0.887 &
    0.089 &
    13.799
\\
\grurecmodel &
    0.034 &
    0.065 &
    0.024 &
    0.252 &
    0.062 &
    11.732 &
    0.042 &
    0.097 &
    0.026 &
    0.365 &
    0.091 &
    11.902
\\
\bertgrurec &
    0.033 &
    0.060 &
    0.025 &
    0.147 &
    0.056 &
   11.200 &
    0.042 &
    0.095 &
    0.028 &
    0.207 &
    0.087 &
   11.405
\\
\bertrec &
    0.033 &
    0.064 &
    0.023 &
    0.108 &
    0.061 &
    11.789 &
    0.041 &
    0.096 &
    0.025 &
    0.161 &
    0.089 &
    11.937
\\
\bertsasrec &
    0.030 &
    0.055 &
    0.022 &
    0.143 &
    0.050 &
   11.096 &
    0.038 &
    0.087 &
    0.024 &
    0.207 &
    0.077 &
   11.307
\\
%\rowcolor{gray!20}
\promptGenItemModel &
    0.032 &
    0.054 &
    0.025 &
    0.548 &
    0.054 &
    13.140 &
    0.037 &
    0.075 &
    0.027 &
    0.728 &
    0.074 &
    13.385
\\
\bertbertrec &
    0.024 &
    0.049 &
    0.017 &
    0.084 &
    0.046 &
   11.219 &
    0.033 &
    0.085 &
    0.019 &
    0.132 &
    0.077 &
   11.442
\\
\grurec &
    0.026 &
    0.051 &
    0.018 &
    0.103 &
    0.046 &
    11.349 &
    0.033 &
    0.080 &
    0.020 &
    0.156 &
    0.073 &
    11.494
\\
\sasrec &
    0.023 &
    0.047 &
    0.016 &
    0.071 &
    0.043 &
    11.081 &
    0.033 &
    0.085 &
    0.018 &
    0.112 &
    0.078 &
    11.316
\\
\promptGenListModel &
    0.029 &
    0.053 &
    0.022 &
    0.703 &
    0.053 &
    13.587 &
    0.033 &
    0.069 &
    0.023 &
    0.831 &
    0.069 &
    13.652
\\
\llamaGenItemModel &
    0.023 &
    0.041 &
    0.018 &
    0.499 &
    0.041 &
    13.366 &
    0.028 &
    0.061 &
    0.019 &
    0.669 &
    0.060 &
    13.451
\\
\srec &
    0.020 &
    0.041 &
    -- &
    -- &
    -- &
    -- &
    0.027 &
    0.067 &
    -- &
    -- &
    -- &
    --
\\
\tallrec &
    0.013 &
    0.024 &
    0.009 &
    0.585 &
    0.024 &
    13.754 &
    0.015 &
    0.035 &
    0.01 &
    0.747 &
    0.034 &
    \textbf{13.824}
\\
%\rowcolor{gray!20}
\promptClassModel &
    0.006 &
    0.012 &
    0.005 &
    0.016 &
    0.010 &
    11.041 &
    0.008 &
    0.017 &
    0.005 &
    0.016 &
    0.012 &
    11.072
\\
MostPopular &
    0.005 &
    0.010 &
    0.003 &
    0.001 &
    0.001 &
    9.187 &
    0.006 &
    0.018 &
    0.003 &
    0.002 &
    0.001 &
    9.408
\\
\hline
\end{tabular}
}
\caption{Evaluation results for the Amazon Beauty dataset}
\label{tab:results-beauty}
\end{table}

\begin{table}[t]
\resizebox{\textwidth}{!}{
\begin{tabular}{lrrrrrr|rrrrrr}
\hline
\multicolumn{1}{c}{} &
\multicolumn{6}{c}{\textbf{Top@10}} &
\multicolumn{6}{c}{\textbf{Top@20}}                                                                   \\
\multicolumn{1}{c}{\multirow{-2}{*}{\textbf{Model}}} &
\textbf{nDCG} &
\textbf{HR} &
\textbf{MRR} &
\textbf{CatCov} &
\textbf{Seren} &
\textbf{Novel} &
\underline{\textbf{nDCG}} &
\textbf{HR} &
\textbf{MRR} &
\textbf{CatCov} &
\textbf{Seren} &
\textbf{Novel}
\\
\hline
\bertmodel &
    \textbf{0.101} &
    \textbf{0.180} &
    0.077 &
    0.240 &
    \textbf{0.151} &
    10.832 &
    \textbf{0.120} &
    \textbf{0.253} &
    \textbf{0.082} &
    0.301 &
    \textbf{0.198} &
    11.028
\\
\popHybrid &
    0.099 &
    0.175 &
    0.075 &
    0.276 &
    0.146 &
    10.823 &
    0.117 &
    0.247 &
    0.080 &
    0.354 &
    0.191 &
    11.021
\\
BERT2BERT4Rec &
    0.098 &
    0.174 &
    0.075 &
    0.208 &
    0.144 &
   10.600 &
    0.116 &
    0.244 &
    0.080 &
    0.281 &
    0.187 &
   10.802
\\
BERT2GRU4Rec &
    0.098 &
    0.175 &
    0.075 &
    0.260 &
    0.147 &
   10.766 &
    0.116 &
    0.245 &
    0.080 &
    0.333 &
    0.189 &
   10.963
\\
\sknnmodel &
0.098 &
0.166 &
\textbf{0.077} &
0.364 &
0.147 &
10.938 &
0.114 &
0.228 &
0.082 &
0.434 &
0.184 &
11.076
\\
\sknn &
0.098 &
0.168 &
0.077 &
0.361 &
0.138 &
10.547 &
0.113 &
0.228 &
0.081 &
0.438 &
0.171 &
10.742
\\
\sasrecmodel &
0.096 &
0.171 &
0.073 &
0.275 &
0.142 &
10.782 &
0.113 &
0.237 &
0.078 &
0.358 &
0.181 &
10.977
\\
\bertrec &
0.095 &
0.169 &
0.0073 &
0.278 &
0.140 &
10.904 &
0.112 &
0.237 &
0.077 &
0.364 &
0.182 &
11.089
\\
\sasrec &
0.087 &
0.153 &
0.067 &
0.153 &
0.126 &
10.784 &
0.103 &
0.217 &
0.071 &
0.197 &
0.163 &
10.997
\\
\grurecmodel &
    0.087 &
    0.154 &
    0.066 &
    0.244 &
    0.127 &
    10.917 &
    0.103 &
    0.218 &
    0.071 &
    0.314 &
    0.163 &
    11.109
\\
\grurec &
    0.086 &
    0.153 &
    0.066 &
    0.232 &
    0.126 &
    10.869 &
    0.103 &
    0.217 &
    0.070 &
    0.304 &
    0.164 &
    11.058
\\
BERT2SASRec &
    0.082 &
    0.149 &
    0.061 &
    0.106 &
    0.115 &
   10.319 &
    0.098 &
    0.212 &
    0.066 &
    0.150 &
    0.152 &
   10.599
\\
\promptGenItemModel &
    0.063 &
    0.116 &
    0.047 &
    0.400 &
    0.107 &
    12.048 &
    0.070 &
    0.144 &
    0.049 &
    0.611 &
    0.123 &
    13.788
\\
\embeddingPop &
    0.051 &
    0.091 &
    0.039 &
    0.270 &
    0.090 &
    13.869 &
    0.060 &
    0.126 &
    0.042 &
    0.369 &
    0.122 &
    13.973
\\
\embeddingModel &
    0.040 &
    0.073 &
    0.030 &
    \textbf{0.684} &
    0.072 &
    \textbf{16.216} &
    0.047 &
    0.102 &
    0.032 &
    \textbf{0.818} &
    0.099 &
    \textbf{16.454}
\\
MostPopular &
    0.024 &
    0.049 &
    0.017 &
    0.000 &
    0.000 &
    7.518 &
    0.032 &
    0.079 &
    0.019 &
    0.001 &
    0.000 &
    7.836
\\
\hline
\end{tabular}
}
\caption{Evaluation results for the \dhr dataset}
\label{tab:results-DH}
\end{table}

\begin{table}[t]
\resizebox{\textwidth}{!}{
\begin{tabular}{lrrrrrr|rrrrrr}
\hline
\multicolumn{1}{c}{} &
\multicolumn{6}{c}{\textbf{Top@10}} &
\multicolumn{6}{c}{\textbf{Top@20}}                                                                   \\
\multicolumn{1}{c}{\multirow{-2}{*}{\textbf{Model}}} &
\textbf{nDCG} &
\textbf{HR} &
\textbf{MRR} &
\textbf{CatCov} &
\textbf{Seren} &
\textbf{Novel} &
\underline{\textbf{nDCG}} &
\textbf{HR} &
\textbf{MRR} &
\textbf{CatCov} &
\textbf{Seren} &
\textbf{Novel}
\\
\hline
%\rowcolor{gray!20}
\sasrecmodel &
\textbf{0.068} &
\textbf{0.129} &
\textbf{0.050} &
0.258 &
\textbf{0.112} &
9.885 &
\textbf{0.085} &
\textbf{0.197} &
\textbf{0.054} &
0.338 &
\textbf{0.150} &
10.026
\\
\bertmodel &
0.068 &
0.128 &
0.049 &
0.266 &
0.110 &
9.662 &
0.084 &
0.194 &
0.054 &
0.345 &
0.143 &
9.862
\\
BERT2BERT4Rec &
    0.067 &
    0.127 &
    0.049 &
    0.238 &
    0.109 &
    9.765 &
    0.083 &
    0.193 &
    0.053 &
    0.313 &
    0.144 &
    9.955
\\
\bertrec &
0.066 &
0.126 &
0.047 &
0.242 &
0.106 &
9.583 &
0.083 &
0.194 &
0.052 &
0.317 &
0.141 &
9.813
\\
\popHybrid &
0.066 &
0.127 &
0.048 &
0.410 &
0.109 &
9.691 &
0.083 &
0.192 &
0.053 &
0.548 &
0.142 &
9.834
\\
\grurecmodel &
0.066 &
0.124 &
0.048 &
0.319 &
0.108 &
9.907 &
0.082 &
0.188 &
0.052 &
0.414 &
0.141 &
10.079
\\
\grurec &
0.065 &
0.123 &
0.048 &
0.389 &
0.106 &
9.976 &
0.081 &
0.188 &
0.052 &
0.482 &
0.141 &
10.161
\\
\sasrec &
0.065 &
0.123 &
0.047 &
0.168 &
0.104 &
9.647 &
0.081 &
0.187 &
0.052 &
0.232 &
0.138 &
9.869
\\
BERT2GRU4Rec &
    0.064 &
    0.122 &
    0.046 &
    0.183 &
    0.105 &
    9.674 &
    0.080 &
    0.186 &
    0.051 &
    0.242 &
    0.137 &
    9.857
\\
BERT2SASRec &
    0.060 &
    0.116 &
    0.043 &
    0.106 &
    0.099 &
    9.527 &
    0.076 &
    0.178 &
    0.048 &
    0.142 &
    0.129 &
    9.738
\\
\srec &
    0.057 &
    0.111 &
    -- &
    -- &
    -- &
    -- &
    0.072 &
    0.173 &
    -- &
    -- &
    -- &
    --
\\
\sknnmodel &
0.051 &
0.099 &
0.037 &
0.243 &
0.072 &
8.789 &
0.064 &
0.151 &
0.041 &
0.353 &
0.093 &
9.053
\\
\ssknn &
0.048 &
0.097 &
0.034 &
0.178 &
0.066 &
8.439 &
0.062 &
0.150 &
0.038 &
0.263 &
0.084 &
8.684
\\
\promptGenItemModel &
    0.030 &
    0.062 &
    0.021 &
    0.457 &
    0.057 &
    10.595 &
    0.037 &
    0.089 &
    0.023 &
    0.686 &
    0.068 &
    11.695
\\
MostPopular &
    0.022 &
    0.042 &
    0.016 &
    0.001 &
    0.000 &
    7.638 &
    0.030 &
    0.075 &
    0.018 &
    0.002 &
    0.001 &
    7.951
\\
\embeddingPop &
    0.024 &
    0.044 &
    0.017 &
    0.328 &
    0.041 &
    11.349 &
    0.030 &
    0.068 &
    0.019 &
    0.483 &
    0.059 &
    11.397
\\
\embeddingModel &
    0.023 &
    0.044 &
    0.017 &
    \textbf{0.656} &
    0.041 &
    \textbf{11.871} &
    0.029 &
    0.067 &
    0.019 &
    \textbf{0.786} &
    0.058 &
    \textbf{11.950}
\\
\llamaGenItemModel &
    0.014 &
    0.025 &
    0.011 &
    0.301 &
    0.022 &
    11.473 &
    0.019 &
    0.045 &
    0.012 &
    0.424 &
    0.031 &
    11.782
\\
\hline
\end{tabular}
}
\caption{Evaluation results for the Steam dataset}
\label{tab:results-steam}
\end{table}

In line with our observations regarding LLM embeddings, BERT embeddings also tend to increase the performance of neural recommendation models, but not consistently for all datasets and all models. In most of the cases, LLM embeddings significantly outperform BERT embeddings except for one case, for the DH dataset, where BERT embeddings elevated GRU4Rec's performance significantly stronger than LLM embeddings.
In summary, the experiments reveal a real potential for leveraging the semantics of a dataset's item descriptions to uplift the performance of the recommendation models. This potential however depends on the dataset and different models capture the semantics in different ways and to different extents. Most of the time, however, embedding initialization leads to significant performance gains especially when LLM embeddings are used.

\tallrec, \srec, and \llamaGenItemModel performed average in the evaluation. Specifically, they trail behind the neural models for the Steam dataset and also behind the LLM generation-based approaches for the Beauty dataset.

Furthermore, \sknnmodel outperforms \sknn on Beauty by 9\% at NDCG@20 and 12\%
at HR@20. However, in the other
datasets, \sknnmodel only slightly improves over \sknn. A notable
observation is that \sknnmodel takes the third place across all
models in the \dhr dataset and exhibits the best MRR@10 metric.
This result is in line with previous
performance comparisons of sequential recommendation models.
In~\cite{Ludewig2018}, for example, it was
observed that the rather simple Sequential Rules (SR) method can lead
to highly competitive results in terms of the MRR for some datasets,
but is not always competitive in terms of Hit Rate. A strong
performance on the MRR measure means that an algorithm is able to
frequently position one positive item in the recommendation lists,
\eg, due to a popularity bias of the algorithm as in the case of the
SR method. Overall, while most accuracy measures are typically highly
correlated~\cite{Anelli2022Top}, some algorithms often perform
particularly well on certain metrics, depending on specific dataset
characteristics. In this respect, our results thus exhibit phenomena
that are similar to those reported in the literature.

The \embeddingPop hybrid closely trailed the top models with a small
uplift in HR@20 compared to \embeddingModel on Beauty. However, for
the other two datasets, it was not competitive, trailing even the
popularity baseline on Steam. That said, besides the tie in NDCG@20 on
Beauty, the \embeddingPop hybrid was consistently better
than \embeddingModel on the other datasets and even achieved a 27\%
increase on the \dhr dataset, showing that the consideration of
popularity can lead to more attractive recommendations. Similar to the
\embeddingPop hybrid, \embeddingModel only exhibited high
performance on the Beauty dataset. In fact, \embeddingModel topped the
MRR metric, and the \embeddingPop hybrid almost matched the % same
performance. Finally, our expectation of high performance on the Steam
dataset
due to the even higher homogeneity of items in a
session than the Beauty dataset according to
Figure~\ref{fig:dataset-embedding-distances}, did not materialize.

Of the fine-tuned models, \promptGenItemModel is the most noteworthy
since it surpasses the performance of \grurec and \sasrec on Beauty,
while \promptGenListModel matches the performance of \grurec and \sasrec on the
same dataset. Note that both models have been fine-tuned on OpenAI GPT
in a single epoch, which was enough for the validation loss to
converge. We highlight a number of notable aspects of the fine-tuned
models in Section~\ref{subsec:model-tuning}. In the remaining
datasets, \promptGenItemModel is by far inferior to mainstream
recommendation models, but also markedly better than simplistic
baselines like MostPopular. Interestingly, the implicit ranking of
recommendations in \promptGenItemModel, where we repeatedly ask the
model to provide single-item recommendations until we have top-$k$,
was more effective than letting \promptGenListModel return a ranked
list of top-$k$ recommendations. This indication is in line with the
performance of \promptRankModel, which although positioned seemingly
high on the performance board, it actually performed worse than
\embeddingModel, which produced the candidate recommendations that
\promptRankModel reranked. Finally, \promptClassModel performed
almost on par with the popularity baseline, showing little to no
potential. On the other hand, we barely experimented with the number
of clusters, which is a crucial performance factor for this task
specification.

\subsubsection{Beyond accuracy metrics}
\label{subsub:beyond-accuracy}

We make the following observations for coverage, serendipity, and
novelty. The \embeddingModel model consistently leads to the best
catalog coverage and novelty except for Novelty@20 in the Beauty dataset where \tallrec performs better. The performance of \embeddingModel in terms of catalog coverage and novelty is not too surprising, given the
nature of the approach, which is solely based on embedding
similarities. Unlike other methods that use collaborative signals,
\ie, past user-item interactions, the general popularity of an item in
terms of the amount of observed past interactions does not play a role
in \embeddingModel, neither directly nor implicitly. Thus, the model
has no tendency to concentrate
the
recommendations on a certain subset of (popular) items.\footnote{We
  recall that the used novelty measure is based on the popularity of
  the items in the recommendations.}
Finally, the serendipity results are aligned with the accuracy measures
across the datasets. This generally confirms the value of
personalizing the recommendations to individual user preferences,
compared to recommending mostly popular items to everyone.\footnote{We iterate
that our serendipity measure counts the fraction of correctly
recommended items that would not be recommended by a popularity-based
approach.}

Interestingly, \sasrecmodel uplifts all beyond-accuracy metrics
compared to \sasrec, especially catalog coverage and serendipity where
it almost doubles catalog coverage on average and contributes an average
of 21\% improvement on serendipity
across datasets. Also notable is the performance of the \popHybrid hybrid,
which is composed of \embeddingModel and \bertmodel, in terms of
catalog coverage. The \popHybrid contributes a 45\% average
improvement in catalog coverage compared to \bertmodel, while
sacrificing only a small margin of performance in terms of accuracy
metrics.
That said, we note that catalog coverage is a coarse metric
that counts an item in the coverage of the item space given just a
single occurrence of the item across all top-$k$ recommendations on the
test set. Still, the improvement is substantial.

\subsection{Observations from Hyperparameter Tuning}
\label{subsec:model-tuning}

Given that our proposed approaches embody a variety of novel technical variations, we devote this section to the relationship between parameters and optimal model performance as revealed by the hyperparameter search process for \embeddingModel (Section~\ref{subsub:hp-embedding}), \seqmodel (Section~\ref{subsub:hp-sequential}), \promptModel (Section~\ref{subsub:hp-fine-tune}), and hybrid models (Section~\ref{subsub:hp-hybrid}). Table~\ref{tab:experiments-map} depicts the number of hyperparameter search trials for each model and dataset.

\subsubsection{Semantic Item Recommendations via LLM Embeddings (\embeddingModel)}
\label{subsub:hp-embedding}

There are four main observations that stand out from the tuning of \embeddingModel across datasets.
First, the use of OpenAI embeddings leads to a clear performance gain compared to embeddings obtained from Google.
Second, dimensionality reduction increases accuracy and, third, the application of PCA always results in the best configuration.
Finally, while reducing the dimensions of the embeddings always helps, the higher the target number of dimensions is, the higher the accuracy.

\subsubsection{LLM-enhanced Sequential Models (\seqmodel)}
\label{subsub:hp-sequential}

For the sequential models enhanced with LLM embeddings, the choice of LLM embedding model per se only leads to marginal performance differences. In fact, both OpenAI and Google embeddings are found in best model configurations depending on the dataset and model. On the other hand, dimensionality reduction consistently contributes to the best model configurations across datasets. Usually, PCA is the optimal reduction method, but LDA and random projection also appear in certain cases. Finally, neural sequential models benefit from low learning and drop rates, while \sknnmodel leverages different types of decay for training and prompt sessions to optimize its performance. \sknnmodel is agnostic to the choice of similarity type, likely due to the normalization of embedding vectors.

\subsubsection{Fine-tuned Models (\promptModel)}
\label{subsub:hp-fine-tune}

A fundamental challenge of fine-tuning regards whether the fine-tuned model will comply with the form of the task at hand. All models that we fine-tuned across datasets closely aligned the form of their responses with the completions provided to them during the fine-tuning process according to the task specifications that we described in Section~\ref{sub:fine-tuned-LLM}.

We iterate that we applied all four \promptModel fine-tuning task
variants on Beauty and only \promptGenItemModel on the \dhr and Steam
datasets due to budget and time limitations.
Section~\ref{subsec:experimental-setup} includes all information
regarding the fine-tuning experiments.
In all setups where both GPT and PaLM are fine-tuned, \ie, for all variants in the Beauty, \dhr, and Steam datasets, GPT outperforms PaLM demonstrating an average uplift of 78\% in terms of NDCG@20.
We include the corresponding results in the online material. For reference, these observations appear to contradict the Chain of Thought Hub~\cite{fu2023chainofthought}, which evaluates the reasoning performance of LLMs, where PaLM 2 exhibits noticeably better performance in a variety of reasoning tasks compared to GPT \texttt{3.5 turbo}.
However, PaLM 2 refers to the \texttt{\small{unicorn}} version, while we used the \texttt{\small{bison}} one, which is smaller than \texttt{\small{unicorn}}.
On the other hand, GPT \texttt{3.5 turbo} performs substantially better than PaLM \texttt{\small{bison}} in the LMSys Chatbot Arena leaderboard,\footnote{\url{https://leaderboard.lmsys.org}} which evaluates the user dialogue with an LLM based on user votes.

In addition, the \promptGenItemModel task has been used to fine-tune
both \texttt{\small{ada}} and \texttt{\small{GPT 3.5 turbo}} OpenAI models on the
Beauty and \dhr datasets. The results are inconclusive. On the one
hand, we observe a notable performance increase of 23\% in terms of
NDCG@20 in favor of \texttt{\small{GPT}} on Beauty, but, surprisingly, on \dhr
\texttt{\small{ada}} performed significantly better than \texttt{\small{GPT}} for the
same task achieving 18\% higher NDCG@20. Given that the comparison
scope is narrow, we can only conclude that the choice of fine-tune
model depends on the dataset and that the evolution of models does not
necessarily lead to higher performance.

With respect to the \textit{temperature} and \textit{top\_p}
hyperparameters, we observe that in all setups a high temperature
results in lower performance as it allows for more creativity at the
cost of item hallucinations. For \promptGenItemModel and
\promptGenListModel a low but non-zero \textit{temperature} leads to
the best performance whereas \promptClassModel and \promptRankModel
perform best with \textit{temperature} set to zero. A possible
explanation for this is that the latter variants do not require any
creativity because the recommendation options are contained in the
provided prompt.

\begin{table}[t]
\rowcolors{1}{}{}
\resizebox{\textwidth}{!}{
\begin{tabular}{lll|rrrrr}
\hline
\textbf{Dataset} &
\textbf{Task} &
\textbf{Model} &
\textbf{HR@20} &
\textbf{NDCG@20} &
\textbf{\# Uniq. Reco.} &
\textbf{Hallucin.} &
\textbf{Emb. Sim.}
\\
\hline
\rowcolor{gray!25} \cellcolor{white} &
\cellcolor{white} &
\cellcolor{white}GPT base &
0.029 &
0.013 &
6553 &
72.38\% &
0.919
\\
&
&
GPT fntd &
0.077 &
0.038 &
20056 &
36.11\% &
0.960
\\
\cline{4-8}
\rowcolor{gray!25}\cellcolor{white} &
\cellcolor{white} &
\cellcolor{white}PaLM base &
0.029 &
0.015 &
 19330 &
78.63\% &
0.944
\\
&
\multirow{-4}{*}{\promptGenItemModel} &
PaLM fntd &
0.043 &
0.020 &
21103 &
56.23\% &
0.950
\\
\cline{4-8}
\rowcolor{gray!25}\cellcolor{white} &
\cellcolor{white} &
\cellcolor{white}GPT base &
0.021 &
0.008   &
28372 &
82.20\% &
0.935 \\
&
&
GPT fntd &
0.069      &
0.033   &
47966 &
43.35\% &
0.953
\\
\cline{4-8}
\rowcolor{gray!25}\cellcolor{white} &
\cellcolor{white} &
\cellcolor{white}PaLM base &
0.022 &
0.009 &
40019 &
78.85\% &
0.940
\\
\multirow{-8}{*}{Beauty} &
\multirow{-4}{*}{\promptGenListModel} &
PaLM fntd &
0.042 &
0.020 &
32643 &
21.24\% &
0.969
\\
\hline
\rowcolor{gray!25}\cellcolor{white} &
\cellcolor{white} &
\cellcolor{white}GPT base &
0.022 &
0.009 &
41951 &
67.39\% &
0.910
\\
&
&
GPT fntd &
0.111 &
0.057  &
55301 &
13.79\% &
0.967
\\
\cline{4-8}
\rowcolor{gray!25}\cellcolor{white} &
\cellcolor{white}&
\cellcolor{white}PaLM base &
0.045 &
0.027     &
50416 &
70.85\% &
0.950
\\
\multirow{-4}{*}{DH} &
\multirow{-4}{*}{\promptGenItemModel} &
PaLM fntd &
0.050 &
0.029  &
43698 &
37.87\% &
0.961
\\
\hline
\rowcolor{gray!25}\cellcolor{white} &
\cellcolor{white} &
\cellcolor{white}GPT base &
0.017
&
0.010
&
2534
&
8.41\%
&
0.858
\\
&
&
GPT fntd &
0.059 &
0.032  &
5701
&
0.36\%
&
0.889
\\
\cline{4-8}
\rowcolor{gray!25}\cellcolor{white} &
\cellcolor{white}&
\cellcolor{white}PaLM base &
0.019 &
0.013 &
13303  &
77.05\% &
0.870
\\
\multirow{-4}{*}{Steam} &
\multirow{-4}{*}{\promptGenItemModel} &
PaLM fntd &
0.025  &
0.014  &
382446 &
58.87\% &
0.867
\\
\hline
\end{tabular}
}
\caption{Comparison between base and fine-tuned models across datasets for
  \promptGenItemModel and \promptGenListModel in terms of HR@20, NDCG@20, number of unique recommendations, percentage of hallucinations, and embedding similarity between hallucinations and closest match.}
\label{tab:base-vs-finetuned}
\end{table}

Finally, to explore the learning capabilities of fine-tuned models, we
compare their performance to the corresponding base models. We observe
that in all setups 
the fine-tuned models significantly outperform their non fine-tuned
counterparts. For the \promptGenItemModel and \promptGenListModel
variants, the model generates recommendations that
are 
not part of the prompt. For those, the base model has a much higher
rate of hallucinations compared to fine-tuned models as
Table~\ref{tab:base-vs-finetuned} shows.
Additionally, the hallucinations produced by the fine-tuned models
exhibit a higher embedding similarity to the catalog items compared to
the base model. These observations indicate that fine-tuned models
recall and prioritize concepts that they processed during their
fine-tuning compared to the base model, which is not given the same
opportunity. Therefore, the base model has to provide recommendations
solely based on the information in the prompt.

For the variants \promptClassModel and \promptRankModel, the
recommendation options are explicitly part of the context such that
the model only has to select and/or re-rank the given item names.
Thus, hallucinations are uncommon in those variants. Still, the
performance advantage of the fine-tuned models over the base models
persists. This indicates that the fine-tuned models learn the
underlying classification and ranking tasks better. This is an expected finding
that is widely acknowledged by literature and practice~\cite{openai2023gpt4, touvron2023llama}.

All in all, the fine-tuned models perform better for each recommendation task with fewer hallucinations and closer semantic proximity to the actual options compared to the base models.
This finding clearly shows that fine-tuning can enable a model to learn not only how to perform a task, but also concepts of a domain to some extent. Thus, it contradicts existing findings and common belief that fine-tuning is only applicable for learning form (instructions) rather than facts~\cite{mitra2023orca-2,kirstain2022examples, ovadia2024finetuning, lewis2021question, wang2023survey, kandpal2023large, luo2023empirical, chen2020recall, Kirkpatrick2017overcoming, berglund2023reversal}.

\subsubsection{Hybrids}
\label{subsub:hp-hybrid}

In both hybrids (\popHybrid, \embeddingPop)
the \embeddingModel model based on the OpenAI embeddings performs better,
while for the \seqmodel\ model performance depends on the dataset:
the OpenAI embeddings dominate
on the \dhr and Steam datasets,
and the Google embeddings on the Beauty dataset.
With regard to dimensionality reduction,
it mainly improves performance compared to the original embedding size.

With respect to the popularity threshold,
for \popHybrid\ we note that increasing the cutoff point seems to decrease performance.
Thus, the hybrid appears to favor \seqmodel\ over \embeddingModel
since a lower threshold results in more recommendations from the former, and
less from the latter.
In contrast,
in \embeddingPop\ increasing the popularity threshold value appears to increase performance,
especially for the \dhr and Steam datasets.

\section{Challenges, Limitations, \& Conclusions}
\label{sec:conclusions}

While large language models offer a variety of new opportunities to build highly effective (sequential) recommender systems, their use, both in industry and academic research, may come with a number of challenges. One central issue in that context is that the most powerful LLMs today are under the control of large organizations and are primarily accessed via APIs. These LLMs are thus black boxes to their users. Moreover, and maybe more importantly, the behavior of these LLMs can change over time and their performance can vary. In an industrial setting, this may represent a major challenge from a quality assurance perspective. In academic settings, the varying behavior of the models can make the exact reproducibility of reported findings almost impossible.

Furthermore, depending on the application use case, an LLM fine-tuning
step may be required to achieve satisfactory recommendation
performance. In cases where the organization that hosts an LLM
provides an %upload
API for fine-tuning, this may lead to data protection and privacy
issues, as potentially sensitive company-internal data may have to be
uploaded for the fine-tuning step. Finally, depending on the
particular way an LLM is leveraged to support (sequential)
recommendation processes, significant costs may arise with the
use of an LLM. Real-world item catalogs can be huge, and fine-tuning
may furthermore be required frequently, \eg, in case of use cases
with highly dynamic item catalogs. In this context, we recall that for
some datasets it may not be sufficient to feed \emph{item names} into
the LLM, as they may carry limited semantic information, as in the
case of the Steam dataset. In such situations, potentially available
longer item descriptions may be used, but this approach may further
add to the cost of the fine-tuning process.

In terms of research limitations, our work so far is based on three
datasets from two domains (e-commerce, games). The datasets used in
our research are however markedly different in terms of various
characteristics (\eg, sparsity, size, available metadata, recurrence patterns, frequency of use) and in terms of their nature (\eg, one long sequence of product reviews per user for Amazon Beauty vs.~a collection of supermarket purchases per user, each consisting of a sequence of products, in the \dhr dataset). Therefore, we are confident that our findings generalize beyond the conducted
experiments. Nonetheless, further experiments are needed to validate
the benefits of leveraging LLMs for sequential recommendation in other
domains, \eg, in music or news recommendation.
In addition, in our current work we have not specifically analyzed the potential of LLM-enhanced sequential recommendations for cold-start situations, where very limited information is available about the preferences of individual users. Also, as another area for future works, it would be interesting to study the use of LLMs for cross-domain recommendation settings.
Finally, future work might include the consideration of alternative LLMs or larger versions of existing LLMs, as these models continue to become more powerful and larger in terms of their parameters. One important research question in that context is if larger model sizes will have a significant impact on recommendation accuracy for the different approaches studied in this paper, see also the analyses in~\cite{Zhang2024Scaling}.

To conclude, Large Language Models and AI assistants like ChatGPT have exhibited disruptive effects in various domains in the past few years. In this work, we have conducted an extensive analysis on the potential benefits of LLMs for the highly relevant class of sequential recommendation problems. Specifically, we have designed, investigated, and systematically assessed different ways of leveraging LLMs in the recommendation process. Our findings show that existing sequential models can particularly benefit from the semantic knowledge encoded in LLMs, leading to substantial accuracy increases compared to previous models. Furthermore, depending on the application, sometimes the LLM-enhanced retrieval of semantically similar items turned out to be highly effective. Our future work includes both the analysis of the proposed methods for additional domains and datasets and the design of alternative ways of building recommendation approaches based on LLMs.

%%
%% The next two lines define the bibliography style to be used, and
%% the bibliography file.
\bibliographystyle{ACM-Reference-Format}
\bibliography{llm-reco}

\end{document}